\documentclass[journal,12pt,onecolumn]{IEEEtran}
\usepackage[utf8]{inputenc}
\usepackage[T1]{fontenc}
\usepackage{lmodern}
\usepackage{hyperref}
\usepackage{amsmath}
\usepackage{graphicx}
\graphicspath{{figures/}}
\usepackage{amssymb} 
\usepackage{multirow}
\usepackage{amsmath}
\usepackage{enumitem}
\newcommand{\vect}[1]{{\mathbf{#1}}}

\usepackage{svn}  
\title{On the economic viability of solar energy when upgrading cellular networks}

 \author{
 \IEEEauthorblockN{Zineb~Garroussi\IEEEauthorrefmark{1}\IEEEauthorrefmark{2}, Abdoul~Wassi~Badirou\IEEEauthorrefmark{1}\IEEEauthorrefmark{2}, Mathieu~D'amours\IEEEauthorrefmark{1}\IEEEauthorrefmark{2}, Andr\'e~Girard\IEEEauthorrefmark{1}\IEEEauthorrefmark{2}, Brunilde~Sans\`o\IEEEauthorrefmark{1}\IEEEauthorrefmark{2}} \\
 \IEEEauthorblockA{\IEEEauthorrefmark{1}Electrical Engineering Department\\Polytechnique Montr\'eal, Qu\'ebec, Canada}  \\
 \IEEEauthorblockA{\IEEEauthorrefmark{2} LORLAB and Research Group in Decision Analysis (GERAD) }  \\
 \{zineb.garroussi, abdoul-wassi.badirou, mathieu.damours, andre.girard, brunilde.sanso\}$@$polymtl.ca}

\begin{document}

\maketitle
\begin{abstract}The massive increase of data traffic, the widespread
  proliferation of wireless applications and the full-scale deployment
  of 5G and the IoT, imply a steep increase in cellular networks
  energy use,  resulting in a significant carbon
  footprint. This paper presents a comprehensive model
  to show the interaction between the networking and energy
  features of the problem and study the economical and technical viability of green networking. 
 Solar equipment, cell
  zooming, energy management and dynamic user allocation are considered in the upgrading network
  planning process. 
  We propose a mixed-integer optimization model to
  minimize long-term capital costs and operational energy expenditures
  in a heterogeneous on-grid cellular network with different types of
  base station, including solar. Based on eight scenarios where
  realistic costs of solar panels, batteries, and inverters were
  considered, we first found that solar base stations are currently
  not economically interesting for cellular operators. We next studied
  the impact of a significant and progressive carbon tax on reducing
  greenhouse gas emissions (GHG). We found that, at current energy and
  equipment prices, a carbon tax ten-fold the current value is the only element that could make
  green base stations economically viable.

\end{abstract}

\section{Introduction}
\label{sec:intro}
Cellular operators are facing traffic growth due to the increasing use
of mobile applications~\cite{bohli19} and the rapid development of
wireless access technologies such as 5G and beyond. To meet this
demand, new base stations (BSs) are being added or upgraded to the
next generation technologies~\cite{chamola16}.


We claim that this current and future upgrading of
cellular systems provides a great opportunity for operators to reduce
their environmental impact. 
The  COP21 Paris agreement has set a target limit of 2°C for the global
warming. 
In that context,  the Information and  Communication Technologies
sector  has a role to play given that it accounts for about 4\% of 
the global energy use~\cite{malmodin18} and has generated approximately
between 1.8\% to 2.8\% of the global GHG emissions in
2020~\cite{freitag21a}. A
significant part of this is produced by  cellular networks~\cite{israr11}
which generate the 
equivalent of 220~MtCO2, which represents 0.4\% of global
emissions~\cite{williams22}.  It is estimated~\cite{wu15b}  that  the
full deployment of 5G could have an
environmental impact  up to~2 to~3 times larger. Moving to
the  millimeter wavebands~\cite{bohli19} will  reduce the range of the
new 5G antennas. This in turn will require a much more dense radio
infrastructure~\cite{ancans17} with the ensuing large increase in
energy use.

One solution is to wait for the grid to use energy sources that don't
emit CO2. This is a major undertaking that will take years to be
implemented. In this paper, we look at a different approach to reduce
the emissions of the wireless networks.
We posit  that it is possible to reduce both OPEX and CO2 emissions
by carefully integrating  the energy management features within the 
planning process while taking into account the CAPEX cost.  

The idea is to make use of
technology that is either currently available or
that will become more common with 5G full deployment: sleep mode, cell zooming, user
allocation and solar power. These can be implemented gradually as the
need arises and do not depend on the general greening of the power
grid. More specifically, we want to see if this evolution can be driven by market economics where the savings provided by the reduction in grid costs are sufficient to cover the additional capital cost of the equipment. In cases where this is not possible, we want to see to what
extent a carbon tax can drive the process.

For this, we propose a model to evaluate the combined technical and economic viability of future
green cellular networks. This is done by a detailed modeling of
energy, communication and demand features and by an in-depth study on
how the issues of energy management, solar energy, CO2 emission, CAPEX
and OPEX are interrelated. 

\section{Literature Review}
\label{sec:review}
The literature relevant to this work is very large, encompassing the
use of solar equipment, energy management, dynamic user assignment and
green planning.
In what follows, we go through a quick technological and literature
survey of each one of those areas.
\subsection{Solar Equipment}
\label{sec:solarequip}
Solar panels have been obvious candidates for green networks for more
than 10~years (See references in~\cite{hu20}). The main advantage of
solar energy is
the fact that the cost per kW has been decreasing steadily over the
last 10~years and is currently the lowest of the more frequently used
green sources~\cite{wikipedia_cost_electricity}.
%
%
Solar panels are
particularly useful and coveted for power stations in remote areas not
covered by the power grid and that use non-renewable resources.
An extreme case is off-grid 
rural areas  where base stations are just diesel
powered.

For instance, the technique proposed in~\cite{alsharif15} decreases
both OPEX and greenhouse gas emissions for remote rural base stations
in Malaysia using solar photovoltaic/diesel generator hybrid power
systems.  The economical and
environmental viability of PV/diesel/battery hybrid system
configuration for a BS in Nigeria was examined
in~\cite{olubayo19}. Due to the large amount of solar 
energy available, this system offers an alternative 
power source for a BS by reducing operational costs and emissions of greenhouse
gases.
A solar PV/Fuel cell hybrid system under
the software HOMER is proposed in~\cite{flavio21}  to power a remote
base station in Ghana. The objective is  to reduce both greenhouse gas
emissions and lower the 
levelized cost of electricity (LCOE). In this hybrid system, the LCOE is
reduced by 67\% compared to diesel power.
Similarly, the authors of~\cite{aderemi18}
use the HOMER software to simulate PV-battery-diesel to power a BS
during 24 hours
under South African climate. They minimize operation costs,
emissions, and power use.
The potential of
photovoltaic to power BS is studied in~\cite{baidas21} for Kuwait
where the
HOMER software is used to determine the 
number of PV, batteries, and converters for an off-grid solar PV system
with diesel generator while minimizing the net present cost in order
to power a cellular BSs.

There have also been some urban implementations.
The work of~\cite{alsharif21b} studies a case in  urban areas of South
Korea where both on- and
off-grid sites use standalone solar batteries to power a macro LTE
cellular base station.
A multi-objective optimization algorithm is proposed
in~\cite{Ibrahim21}  for the  optimal sizing
of a standalone PV/battery to power a BS under the climate of
Sydney. The objective is the minimum annual total life cost.
An optimal decision of demand-side power
management for a green wireless base station is proposed
in~\cite{niyato12} to minimize
the power cost and provide flexibility under traffic load, grid power
price, and renewable source uncertainties.

These studies, rural or urban, all have something in common: they
refer to a single base station and, in most cases, they lack
accounting for the replacement, degradation and installation costs for
the solar equipment.



\subsection{Energy Management and Dynamic User Assignment}
\label{sec:cellzoom}
Given that 80\% of the energy used by a cellular network
comes from the radio infrastructure~\cite{liu14}, radio energy
management must become an integral part of the base station operation.
Even though the existing management
tools are performance-oriented, they make it possible to design
management strategies that can also be
used to reduce energy use, based on the fact that  real-world data
shows that most 
base stations  are underutilized during low traffic periods~\cite{alonso18}.

In fact, it has already been shown that dynamic management may
potentially save between 20\% to 30\% of energy~\cite{boiardi12} in
cellular networks if the system is planned from the start with enough base
stations. These savings are obtained  either by completely turning off
some base stations or by reducing their power during key times of the day. Even
though these methods are nothing new, they will be more easily
implemented in 5G networks and beyond using the notion of \emph{cell zooming}.

%

The actual operation of cell zooming can
be viewed in two ways. In~\cite{mollahasani19}, the total available
power can be reassigned arbitrarily among radio blocks which in turn
can be allocated to different down-link transmissions. It is also
stated in~\cite{jahid20,xu18} that a cell can extend its range through
zooming.
An extreme case of cell zooming is putting some base stations in sleep
mode~\cite{tipper10,budzisz14,vereecken11,wu15}, or even turning them off
completely whenever user demand is lower. Different objective metrics have been
used such as coverage~\cite{wu13}, user demand~\cite{tsilimantos13} or
the actual power usage~\cite{chiaraviglio08b}. A multi-criteria model
is presented in~\cite{niu10} to minimize both
the energy use and the user drop  probability.

A nonlinear model to minimize energy use subject to quality of
service constraints is proposed in~\cite{yigitel14}. The technique
described in~\cite{wang17} uses solar power and on-off operation to
minimize energy 
cost by assigning users to different base stations and choosing solar
or grid power during the day.
 
%
%
One consequence of sleep mode or cell zooming is that one must
re-assign users to different base stations during the day whenever a
user can no longer be served by its base station due to a power 
reduction. This has been examined in detail
in~\cite{boiardi13,boiardi14,damours18}. The conclusion was that
dynamic user assignment can reduce the network cost significantly when
it is integrated into the long-term planning model. These results will
not be repeated here in order to conserve space.  
%
%
%
\subsection{Green Planning}
\label{sec:netmanagissues}
%
The techniques described above can be viewed as network management
and  operate on a short-term horizon of
minutes or hours.
Network planning, on the other hand, where one has to decide on the
installation of base stations and the equipment to go with it, works
on a time scale of many years.
This is by itself a complex procedure since 
one needs to combine cells of different sizes,
including macro, micro, pico, and femto cells.
The size of base stations depends on
several factors, including the site location and the antenna
position. Small stations serve 
areas with high traffic density using a lower amount of
energy~\cite{arnold10,claussen10} but need to be combined with larger
base stations for lower-density areas.

Because of the large difference in time scales, 
the  process was often split  into two independent parts: plan the
equipment upgrades every year and manage the resulting network on a
shorter scale using the equipment available.

Unfortunately, network planning and network management are tightly
coupled because one cannot use a network management technique, such as
solar cells, if the equipment has not been installed.  In fact,  it
has been shown  that integrating
the technology choice with the  long-term planning process produces
networks significantly cheaper than a two-step
procedure~\cite{boiardi13,boiardi14,damours18}. There is thus a
definite need for an \emph{integrated} model that takes into account
the effect of short-term management techniques when deciding on
equipment upgrades.


The problem of minimizing the total CAPEX and OPEX cost over a 10-year
horizon has been examined, albeit for a single base station, 
in~\cite{zhang17}. They make a detailed model of the energy
interactions and study the  trade-off between solar panels, a diesel
generator or grid power.
A similar energy approach was presented in~\cite{wang17} when minimizing
the energy cost of a given network. The variables are the allocation
of users to base stations and the energy source used by each base
station during each one of a given number of time intervals. Even
though both~\cite{zhang17}  and~\cite{wang17} study the energy
planning of the network, neither treats the networking planning
problem, in particular the decision of where and when to locate new
green base stations over the planning horizon.

Some limited work has been done on related
topics. The allocation of users over a
single day to maximize the network operator's revenue has been
examined in~\cite{balakrishnan21}.

The work of~\cite{zheng13} explores the long-term network planning with
green energy harvesting of solar or wind sources. The problem is to
select a subset of candidate BSs and to assign users to the available
base stations subject to  a minimum SINR requirement in order to
minimize the base station installation and the user connection costs
and the cost of electric grid. 


The first attempt at integrating the short-term benefits of solar
energy and network planning has been presented in our previous
work~\cite{damours18}. The goal was to minimize the total operating
plus capital cost of the network.  The users can be re-assigned and
the base station 
antennas can be switched off depending on the demand at different times of day. 


Some work has also considered a marketing-oriented
approach. A financial analysis of network upgrade
is described in~\cite{chen18} to 
optimize the trade-off  between the generated revenue
produced by the upgrade and its cost. A dynamic programming
model is proposed and a fast heuristic solution is used to compute
solutions. A similar approach is used in~\cite{xu15} where a case
study is presented to evaluate the impact of energy trading either
between base stations or through an energy broker and also with
spectrum sharing between operators. 
%
\subsection{Our contribution}
Differently from the above body of work, this paper proposes a
comprehensive approach to study not just the technical or the economic
viability of using solar equipment to update cellular networks, but
the interaction between the two.
The strength of our contribution to the state of the art comes from
a detailed model of energy management and
networking planning. It is precisely this level of detail 
what allows us to clarify the technical and economic
features of network upgrade.
\begin{enumerate}
\item We 
integrate the short-term network and energy management with the long-term network
expansion over  many years into a single model to provide a more
realistic view of how the planning and operation are inter-related.  
\item The planning determines decisions on different base stations
  sizes and types, including solar panels, that are needed to update an
  existing network in an urban area.  
\item The modeling of network and energy operation is very detailed and includes sleep mode, cell zooming,
  user re-allocation, solar, batteries, inverters, controllers to
  reduce the energy use of cellular networks 
\item Very detailed and realistic costs are considered for 
  solar equipment and battery purchase and replacement. We also model
  the degradation of the efficiency of batteries and solar
  equipment   while taking into account
  the time value of money.
\item Realistic evolution of user demand, energy usage and
  illumination profiles are considered as well as regulatory or
  environmental constraints on base station installations. 
\item Co2 emissions and possible taxes are integrated into the
  model.  
\end{enumerate}


We want to use this model to answer some specific questions such as:
\begin{itemize}
\item How large is the cost reduction provided by allowing
  installations over the whole horizon as opposed to a model where the
  installation decisions can be taken only at the beginning of the
  planning horizon?
\item Are the OPEX savings provided by solar energy  large enough to
  justify the added CAPEX?
\item Does cell zooming make solar energy more  economical?
\item Do the benefits of cell zooming add up to those of solar?
\item Is it possible to use cell zooming to reduce the CAPEX by
  delaying the installation of base stations?
\item What is the impact of a carbon tax on the reduction of
 greenhouse gases?
\end{itemize}
\section{Mathematical Model}
\label{sec:mathmod}
As can be seen from the previous discussion, planning a network is a
complex task, from long-term market considerations to technological
choices and real-time network management. 


In this work, we want to see how prices and taxes can be used to
reduce carbon emission using different energy-saving techniques.
For this, we need to \emph{compare} the
effect of these techno-economic options to some base case. If this is to be meaningful,
problems have to be solved to optimality.
Therefore, the model's complexity should not prevent us from finding optimal solutions.
As a result, in the model presented below, we have just kept enough
level of detail for the study to be meaningful.
The approximations and assumptions will be clearly introduced as the model description progresses. 

%
\subsection{Assumptions}
\label{sec:assumptions}
We now briefly review some of the more important simplifications and
assumptions that were made and explain to what extent they are
realistic.
%
%
%
\subsubsection{Network Structure}
\label{sec:netstruct}
In our model, the users are aggregated into so-called \emph{test
  points} which can be viewed as real concentrators or simply as a
collection of users close together.
We are given a set
of available test points for the whole planning horizon, some of which may be
currently inactive. 

The choice of technology is not modeled by separate decision
variables. Instead, we introduce the notion of a base station
\emph{type} which contains a description of the technical features
of the base station, e.g.,  whether it has solar panels or not, its
size, e.g., pico, micro, etc. The test cases can then be run with a
small set of given types.
\subsubsection{Demand}
\label{sec:demand}
Network growth is driven by the increasing number of users and new
applications and services and the quality of service they
require. This involves economic issues such as market forecasting and
technical issues such as power or bandwidth management. 
Because our focus is on energy, we assume that these different kinds
of demands are
converted in an energy requirement from each test point. This can be done by the
engineering or the
traffic department of the operator and is outside the scope of this
paper. A simple example of such a procedure is given in the Appendix.

Demand growth is modeled by activating these test points
at some future time and computing their energy requirement as given by
the demand forecast.
\subsubsection{Forecasting}
\label{sec:forecasting}
In practice, the results of a long-term planning model depend of the
growth forecast provided by the
marketing department. Because
the forecast accuracy decreases for later years, one can use the
model's results for the coming  year
to decide whether  to install new equipment for that year only. The
model can then be run each year with new
forecasts and technology options. This approach is more realistic than
doing a one-year planning since it does take into account the future
demands and technology as they are known at the time when a decision
has to be made.
\subsubsection{Solar  Energy Model}
%
\begin{figure}
  \centering
  \includegraphics[scale=0.5]{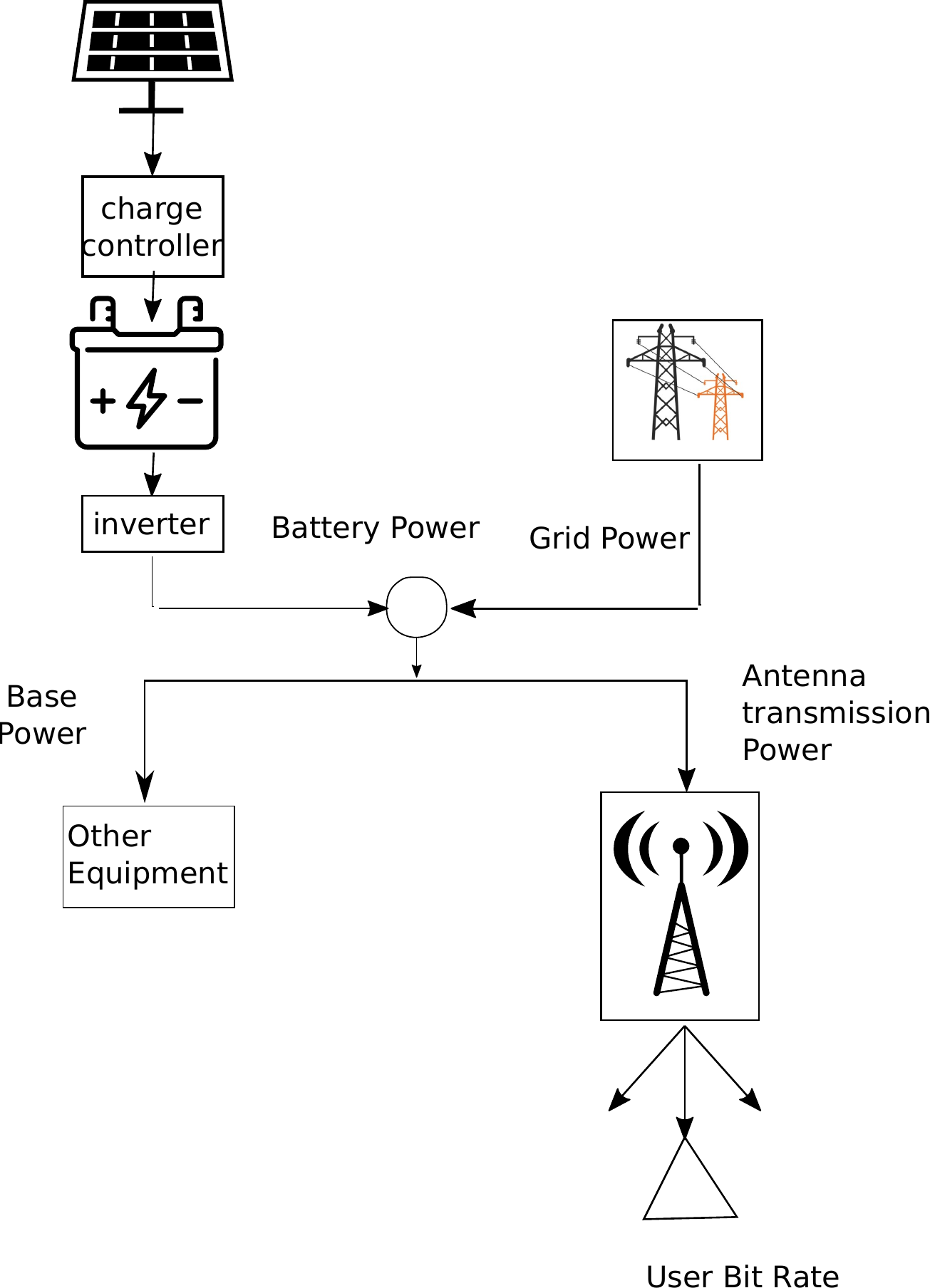}
  \caption{Solar Equipment}
  \label{fig:solarequip}
\end{figure}
An important feature of
solar power is that the energy production can vary widely over short
periods of the order of an hour or less and these variations are
unpredictable.
A minimum amount of solar equipment is thus required to power a green base
station. The main components are  the solar panels and the battery
bank. We also
model the charge controllers  needed to protect the batteries from
overflowing and increase their lifetime. Finally, an AC-DC converter
is also required to power the base station with DC current from its
battery bank. Figure~\ref{fig:solarequip} shows the storage of solar
energy into electrical energy as a back-up energy source to the
electrical grid.

We use a solar profile to 
estimate the amount of electricity produced during the day since replacing the daily variation of sunlight by a
single average value for the whole planning horizon would leave out
the daily changes in sunlight. 
This profile is computed for an \emph{average day} and assume that this is
representative of the operation for the whole year.

This is clearly an
over-simplification since the sunlight available can depend strongly
on the season in regions far from the equator. First note that the
illumination profile is defined for a year and that the actual length
of a ``year'' is a parameter. The model can then easily be used to
take into account the seasonal variations by taking a ``year'' as an
actual season with the corresponding profile. The downside is that
this makes the optimization problem  larger and severely limits the
size of networks that 
can be calculated by an IP solver.

What we have done instead was to
use a single profile corresponding from a region in southern Morocco
with abundant sunlight. Because one of our goals is to
see if solar equipment proves in based on economics alone, this can be
viewed as a most favorable case for solar. Since 
we found that using solar equipment is not
economically justified with this optimistic
profile, the same conclusion would hold under a more realistic profile.
\subsubsection{Cell Zooming}
\label{sec:zooming}
Cell zooming is a power control technique by which a base station can
change its 
transmission power to match certain conditions. Decreasing the power
of a given base station
will decrease its coverage area but may increase that of neighboring ones
by decreasing their SINR. Increasing the power has the opposite effect
of increasing the coverage of the base station and potentially
decreasing the QoS, and hence the coverage of neighbors.
Implementing cell
zooming is a complex issue that requires coordination between base
stations and a knowledge of user demands. This is clearly outside of
the scope of this paper.

Here, we assume that the network can operate
within some specified bounds on the user coverage and QoS when all
base stations transmit at their maximum power. This is the current
mode of operation for networks that don't use power management.
Under this assumption, cell zooming can be modeled by introducing a
state variable for the base 
stations. We assume that a given base station type has a certain
number of power levels that it can use to transmit. The state variable
indicates which of these levels is used at a given time. We also
define a coverage variable that indicates whether a test point is
covered by a base station in some state and add some constraints to
make sure the coverage is logically consistent. The model then computes the
best power levels and the corresponding coverage.

\subsubsection{Costs}
\label{sec:costs}

The objective of the optimization model is to minimize the sum of
capital and operating costs over the planning horizon. While it is
true that an operator would be concerned by the net revenue produced
by the network, in the present case, we assume that the network will always
be able to meet the demand, independently of the decision to use solar
energy or cell zooming. In other words, the revenue is independent of
the choice of technology and can be ignored.
\subsection{Sets}
\label{sec:sets}
Network management has to do with short-term decisions
related to solar power, cell zooming and dynamic user
assignment. These decisions are taken every day on a much shorter time
scale than the planning decisions, typically a few
hours. In this paper, we will use the term \emph{time} to denote them.

The decisions to install base stations is taken on a much larger time
scale, typically a year, and we will use the term \emph{year} to
denote these instants. The sets used in our model are defined in Table~\ref{sets}.
%
\begin{table}[ht]
\begin{tabular}{|l|p{15cm}|}
\hline

$\mathcal{B}$ &  The set of base stations installed at the
  beginning of year~1. This is indexed by  $j$. \\ \hline
$\mathcal{C}$ &  Set of candidate sites where new base
  stations can be installed. We can install at most one base
  station per candidate site. This is indexed by $j$. \\ \hline
$\mathcal{I}$  & Set of test points. A test point
  represents the aggregated demand of a set of users, present
  or future. This is indexed by $i$.  \\ \hline   
    
$\mathcal{L}$ &  Set of base station types. This is
  indexed by $l$. Type $ l = 0$
  corresponds to the base stations in $\mathcal{B}$. \\ \hline

$\mathcal{Q}$  & The set of years that defines the
  planning horizon. This is indexed by $q=1 \dots Q$. The
  traffic demand is defined at the beginning of each year and
  the  planning
  decisions are made at that time.  \\ \hline

$\mathcal{S}_l$ &  Set of states for base stations of type
  $l$. This is currently defined by the transmission power but
  the states could be used to model other operating conditions
  of the base stations. This is indexed by $s=0 \dots S_l$. \\ \hline

$\mathcal{T}$ &  The time instants used to model the daily
  variation of traffic. This is indexed by $t=1 \dots T$.  \\ \hline
  %
\end{tabular} 
\\
\caption{Sets}
\label{sets}  
\end{table}
\subsection{Parameters Values}
\label{sec:paramvals}

\begin{table}[ht]
\begin{tabular}{|l|p{15cm}|}
\hline
$B_{l,q',q}^{+ (-)}$ & Maximum (minimum) amount of energy that
  can be stored in the battery of a base station of type $l$ installed
  in year  $q'$ and used in year $q > q'$. \\ \hline
  
$C_{l,q}$ & The installation cost of a base station of type $l
  \geq 1$ in year  $q$. This is the undiscounted, nominal cost.
This is made up of the construction cost,
  upkeep and software license costs. If the base station has solar
  equipment, this includes the cost of the system used for converting
  solar to electrical energy: solar panels, batteries, converters,
  inverters, etc. \\ \hline
$C^G_{j,q,t}$ & Grid unit energy cost. This is the total charge
  per energy unit charged by the utilities company but excludes any
  emission cost that may be imposed by the regulators.
  We assume in the most
  general case that this can
  depend both on the time period, e.g., time-based grid rates,  
  the base station, if the   supplier of grid power has different
  rates in different regions and that it can change from one year to
  the next. \\ \hline

$E^T_{i,j,q,t}$ & The energy needed by  the station in site $j \in
  \mathcal{B} \cup \mathcal{C}$ to serve
  test point $i$ in period
  $t$ of year  $q$. \\ \hline
$e_{l,q',q,t}^S$ & Amount of electrical energy produced by the
  solar panels in a base station of type $l$ during period $\Delta_t$
  of year  $q$ when these were installed in year $q' < q$. \\ \hline
$k_{i,j,l,s,q,t}$ & Indicator function for coverage. When set
  to~1, indicates that test point $i$ is in the coverage radius of
  base station of type $l$ installed on  site $j \in \mathcal{B} \cup
  \mathcal{C}$ and using power level $W_s, s \in S_l$ during period
  $\Delta_t$ of year $q$. \\ \hline
$m$ & Number of installation periods per year. \\ \hline

$M_{l,j}$ & Indicator function for base station
  installation. When set to~1, this indicates that a base station of
  type $l$ may be installed on candidate site $j$. By definition,
  $M_{0,j\in \mathcal{C}}=0$. \\ \hline
$q_{i}^\ast$ & The first year when test point $i$ becomes
  active. This is used to model the growth of demand. \\ \hline

$r$ & Discount rate. \\ \hline

$U_{l}$ & Indicator function of solar equipment. When set to~1,
  this indicates that a base station of type $l$ has solar
  equipment. By definition, $U_0=0$. \\ \hline

 $W_{l,s}$ & Total power available to  a base station of type $l$ in state
  $s$. By convention, for base stations that use cell zooming,
  $W_{l,0}$ is the power used when idle. For base stations that don't
  use cell zooming, this is the given maximum used power. This
  includes the power used for serving the test points. \\ \hline
$W_{l,s}^{x}$ & Transmission power  used by a base station of
  type $l$ in state $s$.  By convention, for base stations that use
  cell zooming,   $W_{l,0}^x = 0 $. For base stations that don't
  use cell zooming, this is the given maximum transmission power. \\ \hline
$\phi$ & The number of days in an installation period. \\ \hline
$\Delta_t$ & Length of the interval between time instants
  $t$ and $t-1$. \\ \hline
$\lambda_{k,q}$ & Amount of greenhouse gasses emitted per unit
  energy produced, e.g., Mt/kWh,  emitted by a source of type $k$,
  e.g., coal, gas, oil,   etc. \\ \hline
$\pi_q$ &  Unit price per amount of greenhouse gasses emitted,
  e.g, \$/Mt,  in year $q$. \\ \hline  
$\nu_{i,q}$ & Indicator function for test point use. When set
  to~1, indicates that test point $i$ starts to be used. This is
  defined by
  \begin{displaymath}
    \nu_{i,q} = \begin{cases}
      0 & \text{ if $ q < q^\ast_i $}\\
      1 & \text{otherwise.}
    \end{cases}
    \end{displaymath} 
    \\ \hline
\end{tabular}
\\
\caption{Parameters}
\label{parameters}
\end{table}
The parameters are defined in Table~\ref{parameters}. Note that we
need the two year indices  $q$ et $q^\prime$ to describe
the use of solar equipments. For an equipment installed in year
$q^\prime$, the energy production and storage capacity decrease over
time so that they become less efficient as they grow older.
\subsection{Variables}
\label{sec:variables}
In Table~\ref{decision_variables} we describe the decision variables, i.e., the quantities that are
computed by the model. First, a number of decision variables that are
set to~1 when the stated conditions hold.
%

\begin{table}[ht]
\begin{tabular}{|l|p{15cm}|}
\hline
$h_{i,j,q,t}$ &  Test point $i$ is assigned to  site $j
  \in \mathcal{B} \cup \mathcal{C}$  in period $t$ of year
  $q$. \\ \hline

$u_{j,q,t}$ &  Solar batteries are used at site $j \in
  \mathcal{C}$ in period $t$ of year  $q$. \\ \hline

$v_{l,s,j,q,t}$ &  The base station of  type $l$ on site
  $j \in \mathcal{C}$ is in state $s$ in period $t$ of year $q$. \\ \hline
$z_{l,j,q}$ & A base station of type $l$ is installed on
  candidate site  $j \in \mathcal{C}$ in year $q$. \\ \hline
\end{tabular}
\\
\caption{Decision variables}
\label{decision_variables}
\end{table}

We also define some intermediate variables that can be computed from
the parameters and decision variables. These will allow us to simplify
the presentation of the optimization model and they also have a direct
physical interpretation. They are presented in Table~\ref{intermediate_variables}
%
%
\begin{table}[ht]
\begin{tabular}{|l|p{15cm}|}
\hline
$D_{j,q,t}$  & The total amount of energy needed in a  base
  station $j \in \mathcal{B} \cup \mathcal{C}$ in period $t$
  of year $q$ by all test points currently served by the base station. \\ \hline
$\overline{E}_{j,q,t}^P$  & Total amount of energy used by
  base station $j \in \mathcal{B} \cup \mathcal{C}$ during
  period $\Delta_t$ of year $q$ when operating from batteries. \\ \hline
$\overline{E}_{j,q,t}^G$  & Total amount of energy used by
  base station $j \in \mathcal{B} \cup \mathcal{C}$ during
  period $\Delta_t$ of year $q$ when operating from the grid. \\ \hline
$E_{j,q,t}^B$  & Energy available from the batteries of
  base  station $j \in \mathcal{C}$ at the beginning of period
  $t$ of year $q$. \\ \hline
$E_{j,q,t}^S$ & Energy produced by the solar panels of
  base station  $j\in \mathcal{C}$ during period $t$ of year
  $q$. \\ \hline
  $L_{j,q,t}$  & Solar energy lost at base station  $j\in
  \mathcal{C}$ during period $t$ of year $q$. \\ \hline
  $w_{l,j,q}$  & Indicator variable set to~1 if a base station
  of  type $l$ is currently installed on site $j \in
  \mathcal{B} \cup \mathcal{C}$ in year  $q$. This can be seen
  as the indicator of the state of site $j \in \mathcal{B}
  \cup \mathcal{C}$ in year  $q$. \\ \hline

\end{tabular}
\\
\caption{Intermediate variables 1}
\label{intermediate_variables}
\end{table}

We also need some intermediate variables to take into account the
nonlinear terms that will show up in the model, they are defined in
Table~\ref{intermediate_variables2}.
%

\begin{table}[ht]
\begin{tabular}{|l|p{15cm}|}
\hline	
$x_{l,s,j,q,t}$ & Binary variable set to~1 if the base
          station of  type $l$ that is installed at site $j \in
          \mathcal{C}$ operates in state $s$ using the batteries
          during period $t$ of year $q$. These variables are given by
          $x_{l,s,j,q,t}=v_{l,s,j,q,t}u_{j,q,t}$. \\ \hline
\end{tabular}
\\
\caption{Intermediate variables 2}
\label{intermediate_variables2}
\end{table}
\subsection{Objective Function}
\label{sec:objfunc}
%

The capital cost
depends on the type of installed base stations and the installation
year. Operating cost come from the grid cost for operating the base
stations and the emission costs. Installing solar equipment
can then allow a reduction in the 
operating cost, as well as a reduction of greenhouse gases, but at the
expense of an increase in the capital cost.
The objective function can then be written as:
 \begin{eqnarray} 
 \tiny
 Z = \sum_{q \in \mathcal{Q}} (1 + r)^{-q }  \sum_{j \in \mathcal{C}}   \sum_{l \in \mathcal{L}} C_{l,j,q} z_{l,j,q} \nonumber \\
      + 
    \sum_{q \in \mathcal{Q}} (1 + r)^{-q }      m \phi  \sum_{j\in \mathcal{B} \cup \mathcal{C}} \sum_{t \in \mathcal{T}} \left\{   C^G_{j,t} + \pi_q \lambda_{k,q}
        \right\} \overline{E}^G_{j,q,t}                
 \label{eq:objectif}.
 \end{eqnarray}
\subsection{Installation of Base Stations}
\label{sec:insbs}
We now describe the various constraints that may limit the
installation of certain base stations in some sites.

First, some base station types might not be allowed on some sites
because of physical or environmental conditions. This yield the constraint
\begin{equation}
  \label{eq:inst1}
\sum_{q\in \mathcal{Q}} z_{l,j,q} \leq M_{l,j} \qquad \forall l \in \mathcal{L} , j \in \mathcal{C}.
\end{equation}
Given that $M_{0,j\in \mathcal{C}}=0$, then if follows that
$z_{0,j,q}=0 \quad \forall j \in \mathcal{C}, q \in \mathcal{Q}$. 
There is no replacement of base stations during the planning horizon
so that a base station can be installed on a given site at
most once, which yields the constraint
\begin{equation}
  \label{eq:inst2}
\sum_{l\in \mathcal{L},q\in \mathcal{Q}} z_{l,j,q} \leq 1 \qquad \forall j\in \mathcal{C}.
\end{equation}
If a base station is currently installed on some site at a given time,
it must have been installed first at some previous time, so that
\begin{align}
  \label{eq:defwz}
%
  w_{l,j,q} =\sum_{q'\leq q} z_{l,j,q'} \qquad \forall l \in \mathcal{L}, j \in \mathcal{C}, q\in \mathcal{Q} .
\end{align}
In order for a base station to use solar energy, the type must have
solar capability
\begin{equation}
  \label{eq:solarcap}
  u_{j,q,t} \leq \sum_{l\in \mathcal{L}} w_{l,j,q} U_l \qquad \forall j\in \mathcal{C}, q \in \mathcal{Q}, t \in \mathcal{T}.
\end{equation}
\subsection{Test Point Assignment}
\label{sec:tpassign}
All currently used test points must be assigned to a site:
\begin{equation}
  \label{eq:assignalltp}
  \sum_{j\in \mathcal{B} \cup \mathcal{C}} h_{i,j,q,t} = \nu_{i,q} \qquad \forall i\in \mathcal{I}, q \in \mathcal{Q}, t \in \mathcal{T}.
\end{equation}
The assignment is possible only if the coverage radius of the base
station is large enough so that
\begin{align} 
  h_{i,j,q,t} &\leq \sum_{\substack{l\in \mathcal{L} \geq 1 \\
      s \in   \mathcal{S}_l}} v_{l,s,j,q,t} k_{i,j,l,s,q,t} &  \nonumber \\
      & \forall i\in
  \mathcal{I}, j \in
  \mathcal{C}, q \in
  \mathcal{Q}, t \in
  \mathcal{T}
  \label{eq:assigncov}
  \\
  h_{i,j,q,t} &\leq \sum_{\substack{l\in \mathcal{L} \nonumber \\
      s \in   \mathcal{S}_l}} k_{i,j,l,s,q,t} & \\       
  &    \forall i\in \mathcal{I}, j \in
  \mathcal{B} \cup \mathcal{C}, q
  \in \mathcal{Q}, t \in \mathcal{T}
  \label{eq:assigntpcovexist}.
\end{align}
\subsection{Energy Storage}
\label{sec:bscap}
The amount of energy available in a base station must be large enough
to meet the demand for the test points that it serves
\begin{align}
  D_{j,q,t}&=\sum_{i \in \mathcal{I}} E_{i,j,q,t}^T h_{i,j,q,t} & \nonumber \\
 & \forall   j \in   \mathcal{B}   \cup   \mathcal{C}, 
  q \in   \mathcal{Q},   t \in   \mathcal{T}
  \label{eq:demanddh}
  \\
  D_{j,q,t} &\leq \Delta_t W_{0,0}^{x} &  \nonumber \\ 
  & \forall  j \in \mathcal{B}, q
  \in \mathcal{Q}, t \in   \mathcal{T}
  \label{eq:demandinstsite}
  \\
  D_{j,q,t} &\leq \Delta_t \sum_{l \in \mathcal{L} \geq 1, s \in
    \mathcal{S}_l} v_{l,s,j,q,t} W_{l,s}^{x} &\forall  j \in
  \mathcal{C}, q   \in   \mathcal{Q}, t   \in \mathcal{T}
  \label{eq:demandcandsite}.
\end{align}
\subsection{Energy Production Model}
\label{sec:energymod}
The base station installed on some site $j$ must be in a single state
for every period $t$ of year $q$. This is modelled by  variable
$w_{l,j,q}$. If a station of  type $l$ has not yet been installed on
site $j$ in year $q$, then $w_{l,j,q}=0 \Rightarrow v_{l,s,j,q,t}=0
\quad \forall s \in \mathcal{S}_l$.
On the other hand, if type $l$ has already been installed, then
$w_{l,j,q}=1$ which forces the selection of a single transmission power.
\begin{equation}
  \label{eq:singlepower}
\sum_{s \in \mathcal{S}_l} v_{l,s,j,q,t} = w_{l,j,q} \qquad \forall l\in \mathcal{L}, j \in \mathcal{C}, q \in \mathcal{Q}, t \in \mathcal{T}.
\end{equation}

A base station must operate either in battery or grid mode but not
both. This makes the model somewhat more complex. 
If we power the antennas from the batteries, we have to meet two
conditons: the base station type that is currently installed must be
of the right type and we must have decided to use battery power during
that time. This yields:
\begin{align} 
  \overline{E}_{j,q,t}^P &=  \Delta_t \sum_{l \in \mathcal{L} \geq 1, s
    \in \mathcal{S}_l} W_{l,s} v_{l,s,j,q,t}
  u_{j,q,t} \nonumber \\
  & = \Delta_t \sum_{l \in \mathcal{L} \geq 1, s
    \in \mathcal{S}_l} W_{l,s} x_{l,s,j,q,t} ~~~ \forall  j \in \mathcal{C}, q \in \mathcal{Q}, t \in \mathcal{T}
  &
  \label{eq:linbattmode}
  \\
  \overline{E}_{j,q,t}^P &=0~~~\forall  j \in \mathcal{B}, q \in
  \mathcal{Q}, t \in \mathcal{T} & \label{eq:energie1} .
\end{align}

If we power from the grid, the conditions are that the right type of
base station is installed and we have decided not to use battery
power, i.e., 
\begin{align} 
  \overline{E}^G_{j,q,t} &= \Delta_t \sum_{l \in \mathcal{L} \geq 1, s
    \in \mathcal{S}_l} W_{l,s} v_{l,s,j,q,t} ( 1 -  u_{j,q,t})
  \nonumber \\
  & = \Delta_t \sum_{l \in \mathcal{L} \geq 1, s
    \in \mathcal{S}_l} W_{l,s} (v_{l,s,j,q,t}-  x_{l,s,j,q,t})
  & \nonumber \\ 
 & \forall  j \in \mathcal{C},   q \in \mathcal{Q},   t \in
  \mathcal{T}
  \label{eq:energygridlin}
  \\
  \overline{E}_{j,q,t}^G &=\Delta_t W_{0,0}
  & \nonumber \\ 
  & ~~\forall  j \in \mathcal{B},   q \in \mathcal{Q}, t \in
  \mathcal{T}
  \label{eq:energygridinst}.
\end{align}
Note that both~\eqref{eq:linbattmode} and~\eqref{eq:energygridinst} introduce nonlinear terms in the 
model. This is taken care of in section~\ref{eq:lineariz}.

The energy produced by the solar equipments is defined by:
\begin{equation}
  \label{eq:solenerg}
%
  {E}^S_{j,q,t} = \sum_{l \in \mathcal{L}, q' \in \mathcal{Q}} e_{l,q',q,t}^S z_{l,j,q'} \qquad   \nonumber \\ ~~~ \forall  j \in \mathcal{C}, q \in \mathcal{Q}, t \in \mathcal{T} .
\end{equation}
The energy available at the beginning of period $t$ is given by:
\begin{align}
  E^B_{j,q,t} &= E^B_{j,q,t-1}  + E^S_{j,q,t-1} - L_{j,q,t-1} -
  \overline{E}^P_{j,q,t-1} & \nonumber \\ 
 &  ~~~ \forall  j \in \mathcal{C}, q
  \in \mathcal{Q}, t \in
  \mathcal{T} \geq 2
  \label{eq:battenergy}
  \\
  E^B_{j,q,1} &= E^B_{j,q,T}  + E^S_{j,q,T} - L_{j,q,T} -
  \overline{E}^P_{j,q,T} & \nonumber \\ 
 &  ~~~ \forall  j \in \mathcal{C}, q
  \in \mathcal{Q}
  \label{eq:battenergy0}
\end{align}
The storage capacity of batteries is bounded above and below by:
\begin{align}
  \label{eq:battbnd}
%
\sum_{l \in \mathcal{L}, q' \in \mathcal{Q}} z_{l,j,q'} B^-_{l,q',q} \leq E^B_{j,q,t} \leq \sum_{l \in \mathcal{L},q' \in \mathcal{Q}} z_{l,j,q'} B_{l,q',q}^+ \qquad & \nonumber \\ 
   ~~~ \forall  j \in \mathcal{C}, q \in \mathcal{Q}, t \in \mathcal{T} 
\end{align}
In a given period, a base station with solar equipment cannot lose more than the solar energy produced by the panels during this period:
\begin{equation}
  \label{eq:losseq}
  %
  0 \leq L_{j,q,t} \leq E^S_{j,q,t} \qquad \forall  j \in \mathcal{C}, q \in \mathcal{Q}, t \in \mathcal{T} 
\end{equation}
\subsection{Linearization}
\label{eq:lineariz}
As written, the model contains nonlinear constraints  due to the
presence of the term $v_{l,s,j,q,t}u_{j,q,t}$
in~\eqref{eq:linbattmode} and~\eqref{eq:energygridinst}. We can make 
the constraints linear by adding a set of independent variables
$x_{l,s,j,q,t}=v_{l,s,j,q,t}u_{j,q,t}$ and adding the constraints:
\begin{align}
  x_{l,s,j,q,t} &\leq v_{l,s,j,q,t} \label{eq:linearisation1}
  \\
  x_{l,s,j,q,t} &\leq u_{j,q,t} \label{eq:linearisation2}
  \\
  x_{l,s,j,q,t} &\geq v_{l,s,j,q,t} + u_{j,q,t} -1 \label{eq:linearisation3}
  \\
  \forall l \in \mathcal{L}, s \in \mathcal{S}_l,j &\in \mathcal{C}, q \in \mathcal{Q}, t \in \mathcal{T} \nonumber
\end{align}
\section{Data Sets}
\label{eq:datasets}
In order for the results to be meaningful, we tried as best as
possible to use realistic data for the costs and other technical
parameters.
The cost of a base station is made up of the building and installation
cost in addition to the cost of the solar equipment. This is given by:
\begin{align}
  C_{l,q} = C^{B}_l + C^S_q 
  \label{eq:bscost}
\end{align}
where
\begin{description}
\item $C^{B}_l$ Construction and installation cost of a base
  station of type $l$~\cite{johansson04},
\item $C^S_q$ Purchase and installation cost of the solar
  equipment. This depends of the maximum power that can be
  generated and the battery storage cost. According
  to~\cite{eia-solar-cost21}, a realistic value is
  3\$/W. This cost is indexed by year to model the expected
  decrease of the cost of solar generation.
\end{description}
\begin{figure}
  \centering
  \includegraphics[scale=0.5]{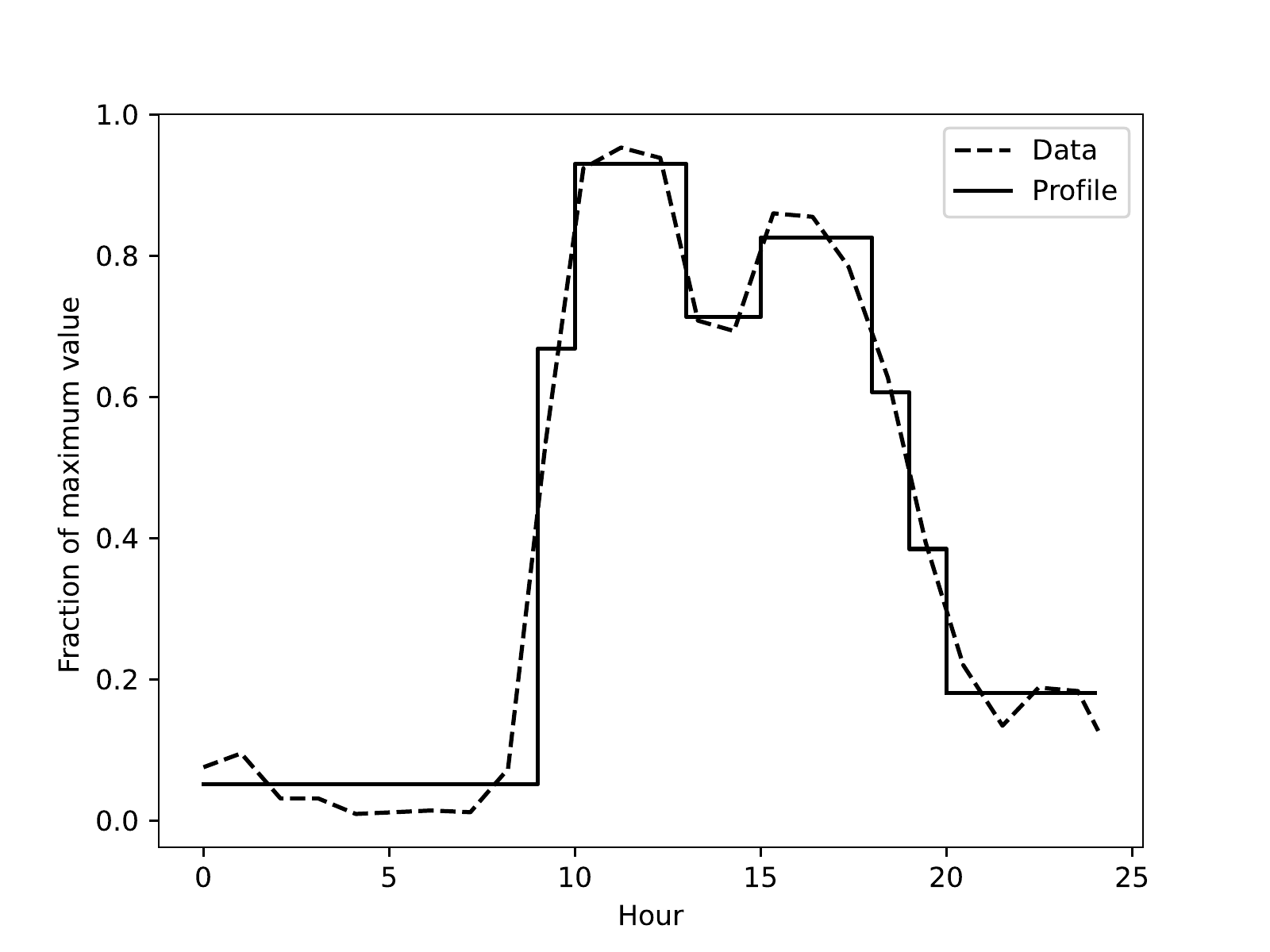}
  \caption{Traffic profile}
  \label{fig:trafprofile}
\end{figure}
\begin{figure}
  \centering
  \includegraphics[scale=0.5]{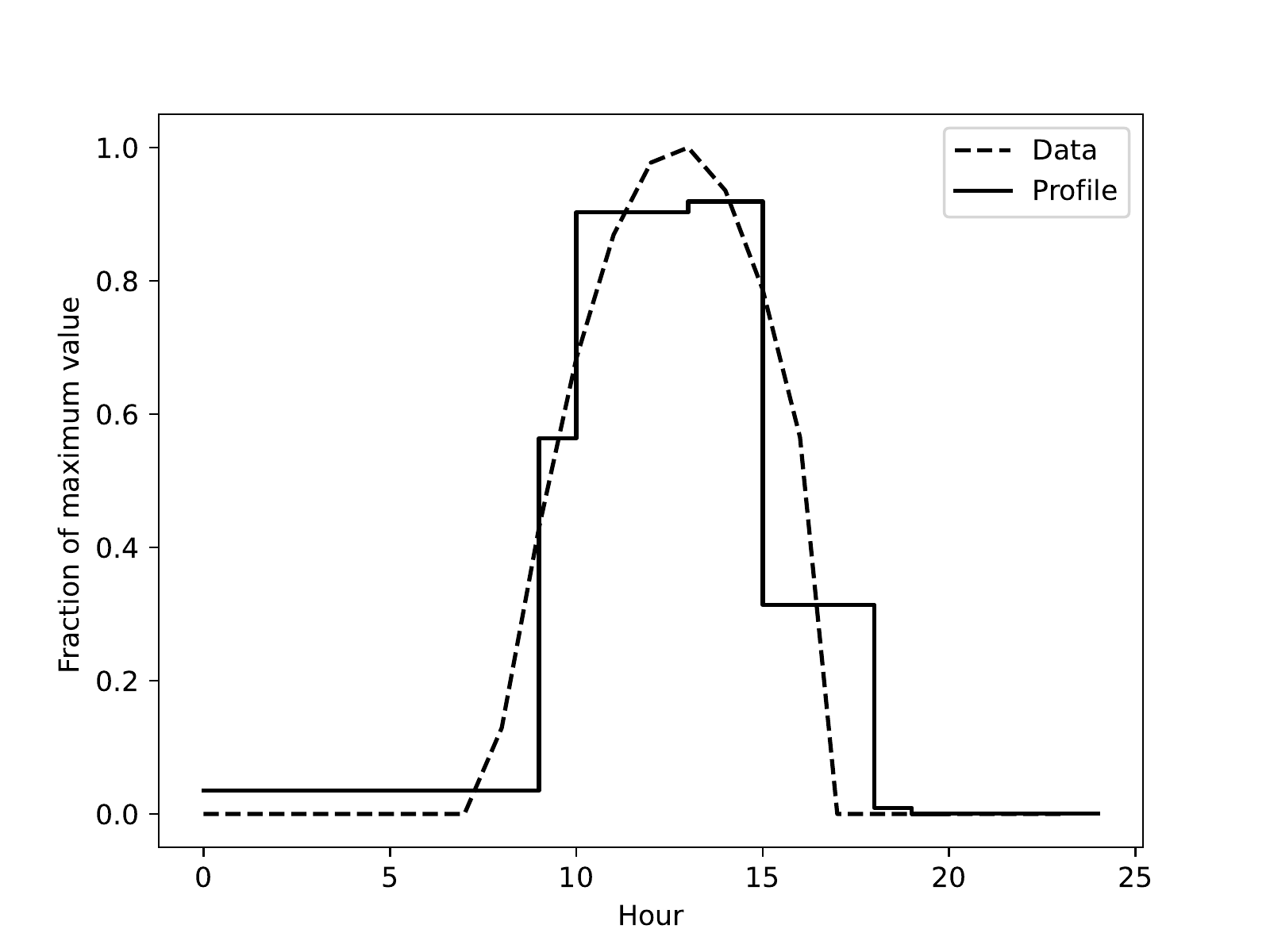}
  \caption{Illumination profile}
  \label{fig:illuprofile}
\end{figure}
The numerical results were computed for two small networks P1 and P2
with different peak traffic profiles. The parameters common to
all cases are shown in Table~\ref{tab:commonparams}. We have split the
day into eight different periods to get a more accurate view of the
changes in the user demands. These are shown on
Figure~\ref{fig:trafprofile}. They correspond to a workday profile in
an inner core cell~\cite{marsan13}. 

The initial peak user rate for P1 is 10Mbps and grows at a~20\% yearly
rate so that user demand doubles every four years. The peak user rate
for P2 is 12Mbps. We have used a~12\%  discount rate and a~2.64\%
inflation rate which reduces by half the value of a
capital expenditure after~6 years. A unit cost of
\$0.20/kWh is representtive of the power rates in many countries. In
Canada, for instance, this varies~\cite{energyhub21} from 
\$0.07/kWh to  \$0.38/kWh for residential users. The cost of solar
equipment is critical for the introduction of solar energy in
networks. In northern countries, a realistic
value~\cite{HydroQuebecInstaller} of \$3/W is a threshold for having
profitable solar power. In the numerical work, we have used the same
cost over the whole planning horizon.  

An antenna gain of~3 is typical for current antennas. The  propagation
coefficient and the channel noise are typical values for urban
networks. Figure~\ref{fig:illuprofile} shows the illumination profile
for a city in southern Morocco~\cite{ouarzazate22}.
The peak illumination is set to 800 $W/m^2$.  

\begin{table}[h]
  \centering
  \begin{tabular}{rc}
    Parameter & Value \\  \hline
    No periods & 8\\
    Discount rate & 0.12 \\ 
    Inflation rate & 0.0264 \\     
    Annual demand growth rate & 0.20 \\
    Unit cost of grid energy(\$/kWh) & 0.20 \\
    Unit cost of solar equipment (\$/w) & 3 \\
    Antenna gain & 3 \\   \\
  \end{tabular}
  \caption{Common parameters for all networks}
  \label{tab:commonparams}
\end{table}

\begin{table}
  \centering
  
  \begin{tabular}{llllllll}
    Pb No & No years & No Macro  BS & No CS & No TP & {Maximum initial user rate (Mbps)} & Channel noise & Propagation coefficient \\
    \hline
    P1	& 10 & 1  & 8  & 18 & 10 & $10^{-5}$ & 3 \\
    P2	& 10 & 1  & 14 & 30 & 12 & $1.5e-5$  & 2 \\  \hline\\
  \end{tabular} 

  \caption{Specific network parameters\\
    BS: Base Station, CS: Candidate Sites, TP: Test Points}
  \label{tab:specificparams}
\end{table}

The parameters that vary for each network are shown on
Table~\ref{tab:specificparams}. They define the size of each
problem. All cases are for a  10-year horizon. This value was
chosen as a compromise between having a problem size small enough to
be able to solve it in reasonable time and having a large enough
horizon to allow significant discounting and installation delays.

The macro base stations are the ones currently in the operator's
network.  We use only one of these so that
the candidate sites will be needed to meet the demand.
We limit the study to small networks because we want to compare the
total cost for different scenarios. For this to be meaningful, we need
to compute optimal solutions or at least solutions with a small gap. 
This can be done in  a reasonable execution time only for small networks.
\section{Scenarios}
The  options that are compared are summarized in a number of
scenarios.
\paragraph{ B: Base Network}
\label{sec:scp1}
This is  the reference network. It  has no solar equipment, no
cell zooming and no dynamic management. It corresponds to most of the
base stations currently used in 4G networks.  
\paragraph{ S: Solar Power}
\label{sec:scp2}
In this scenario, only solar is enabled so that we can  quantify the
savings from solar power. One goal is to try to determine a
relationship between the unit cost of 
solar power on the one hand and the capital cost of the solar
equipment and the grid cost on the other. We also want to see, in the
case where the unit cost of solar is smaller than the grid cost,
whether the best solution is to install solar equipment everywhere.  
\paragraph{ O: On-OFF}
\label{sec:scp3}
In this scenario, only the BS sleeping mode is enabled, when a
base station can be put on idle power during a low traffic
period. This is a special case of cell zooming. 
\paragraph{ Z: Cell Zooming}
\label{sec:scp4}
Here,  only cell zooming is enabled. With cell zooming, we can increase the
reach of  base stations in a given direction while reducing it
elsewhere. We can then evaluate the impact of cell zooming on the network
cost. This will be reflected by a reduction of the operating costs and
may have an impact on the capital cost.
\paragraph{ S+O: Solar and On-off}
\label{sec:scp5}
This  is a combination between scenario S and scenario On-OFF where
both technologies are used at the same time. This will allow us to see
whether the benefits of the two technologies add up.
 \paragraph{ S+Z: Solar and Cell Zooming}
\label{sec:scp6}
In this scenario, both solar and cell zooming are
available. This is the most important one, because it
optimizes the dynamic management and the use of the solar together to
achieve a truly optimal solution.
\paragraph{S + Z0}
\label{sec:scp7}
Here, both solar and cell zooming are available but can be installed
only in the first year. This corresponds to a planning technique that
does not consider the time evolution of the network, as was done
in~\cite{damours18}.
\paragraph{FS + Z}
\label{sec:scp8}
This scenario is used to compare a solution where we force the
installation of solar equipment on all BS. That case is 
interesting to know if a purely solar solution gives reduced costs and
if not, how sub-optimal it is.

These scenarios are implemented in the model by fixing different variables at zero
while leaving the others free.
This is summarized in Table~\ref{state_variables_scenarios}.  For 
simplicity and brevity, we only illustrate solutions to the
peak period the last year which corresponds to
the peak of demand for TP on the planning horizon. Any feasible
solution at this time is also feasible for all other periods.

\begin{table}
\centering
\begin{tabular}{llllllll}  \hline

  Scnr & S & O & Z & $z$ & $v$ & $u$ & $h$ \\
\hline
B & no & no & no & F & 0 & 0 & F \\
S & yes & no & no & F & 0 & F & F \\
O & no & yes & no & F & *** & 0 & F \\
Z & no & yes & yes & F & 0 & 0 & F \\
S+O & yes & yes & no & F & *** & F & F \\
S+Z & yes & yes & yes & F & 0 & F & F \\
S+Z0 & yes & yes & yes & * & 0 & F & F \\
FS+Z & yes & yes & yes & ** & F & F & F \\


\end{tabular}
\caption{State of variables for each Scenario: *: Free in first year
only **: Only solar types allowed ***: Two states only. F: variable is
free}
\label{state_variables_scenarios}
\end{table}
\section{Numerical Results}
\label{sec:numres}
%
%
%
%
%
In this part, we analyze the results from the optimization
model with and without GHG taxes. We use two small networks, P1 and
P2,  generated
from the same data except for channel noise, propagation coefficient
and the demand with  low values for network P1 and high for P2.

In order to produce optimal solutions, the problem and its variants
are modeled by AMPL and solved using CPLEX with the default
options. AMPL integrates pre-solving mechanisms which allow certain
variables and constraints to be eliminated before calling on the
solver. We used a standard IP solver because we must have an exact solution
for all scenarios
in order to compare the costs  of solar and sleep modes.  Larger networks
will require heuristic techniques specifically adapted to the
optimization model. The algorithms were executed on Intel(R) Core(TM) i7-8700 CPU $@$ 3.20GHz with 12 cores running at 3.5 GHz and 65 GBs RAM.

Tables~\ref{Results without taxes (Costs)} and~\ref{Results
  without taxes (Energy)} give various costs and amounts of energy  for
the two networks without  GHG taxes.
The column titles represent
  \begin{description}
  \item[$Z$] Total network cost. This is the sum of $Z_c$ and $Z_{op}$
  \item[$\Delta$] Change in $Z$ relative to the cost of scenario B
  \item[$Z_c$] Total capital cost including solar equipment
  \item[$Z_s$] Capital cost of solar equipment
  \item[$Z_{op}$] Operating cost
  \item[$Z_g$] Cost of grid energy
  \item[$Z_{CO_2}$] Cost of greenhouse taxes
  \item[$Z_s$/kWh] Production cost of solar energy = $Z_s$/(Total
    energy produced by the panels)
  \item[$Z_s^\ast$/kWh] Production cost for the energy 
    used. This is 
     $Z_s$/(Total energy used by the network)
  \end{description}


The second set of  tables~\ref{Results without taxes (Energy)}
and~\ref{Results with taxes (Energy)} shows energy information where all energy values are in MWh.
The column labels are defined as 
\begin{description}
    \item[$E_N$]  This is the total amount of energy
      that the network needs
    \item[$E_G$]  The amount of  energy coming from the grid
    \item[CO$_2$]  The number of tons of CO2 produced by the network
    \item[$E_{si}$]   The amount of
      energy produced by the installed solar equipment
    \item[$E_{su}$] This is the amount of solar energy effectively used
\end{description}
%
%
The difference $E_{si} - E_{su}$ is the energy lost due to  the limited
capacity of batteries. 
The $Z_s$/kWh
column gives the cost in dollars per kilowatt-hour of solar
based on the amount of solar energy actually used. This is simply
the value of the CAPEX column divided by the solar energy used. The $Z_s^\ast$/kWh
column is the energy lost due to the
limited capacity batteries. Solar losses are the ratio $(E_{si} -
E_{su})/E_{si}$.  This ratio clearly shows that
installing too much solar equipment will increase losses and
installing larger batteries is too expensive.
\begin{figure*}
\includegraphics[width=\linewidth,height=5cm]{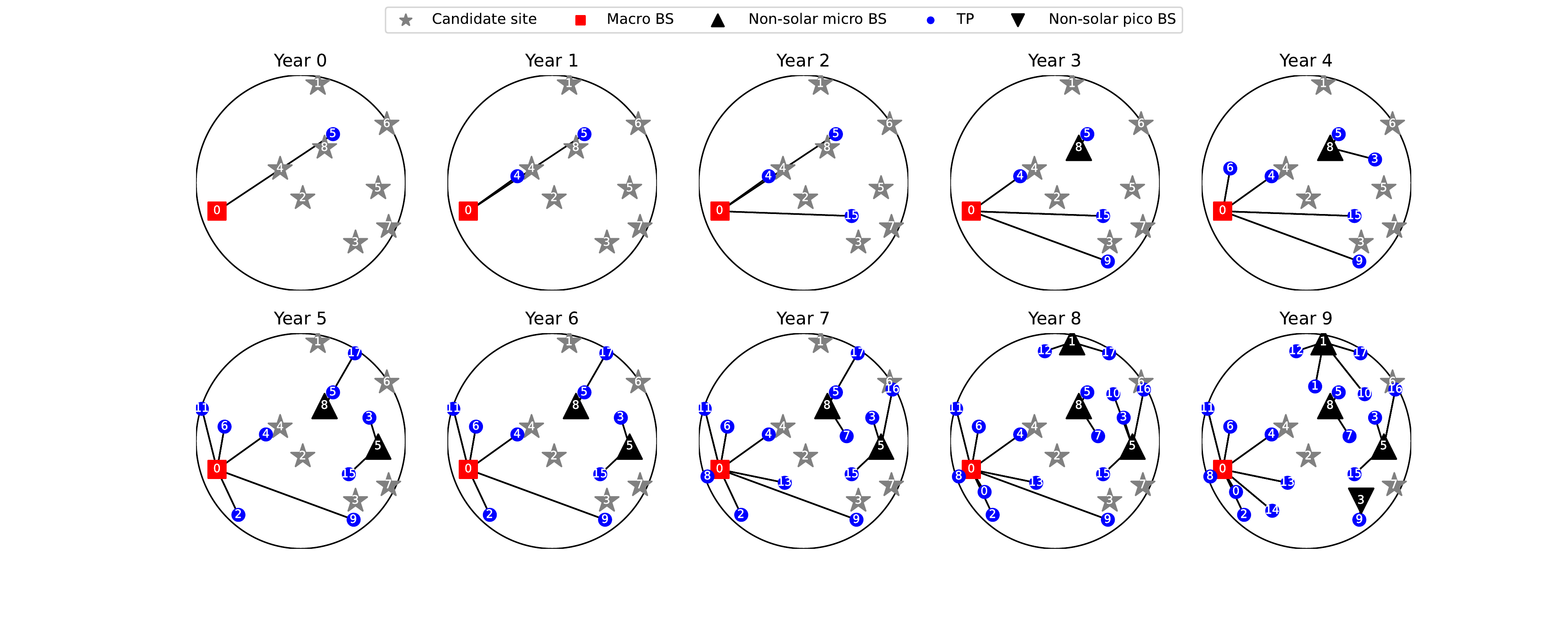}
\caption{Network P1 (Without taxes and Scenario S+Z)}
\label{Network P1(Without taxes and Scenario S+Z)}
\end{figure*}

\begin{figure*}
\includegraphics[width=\linewidth,height=5cm]{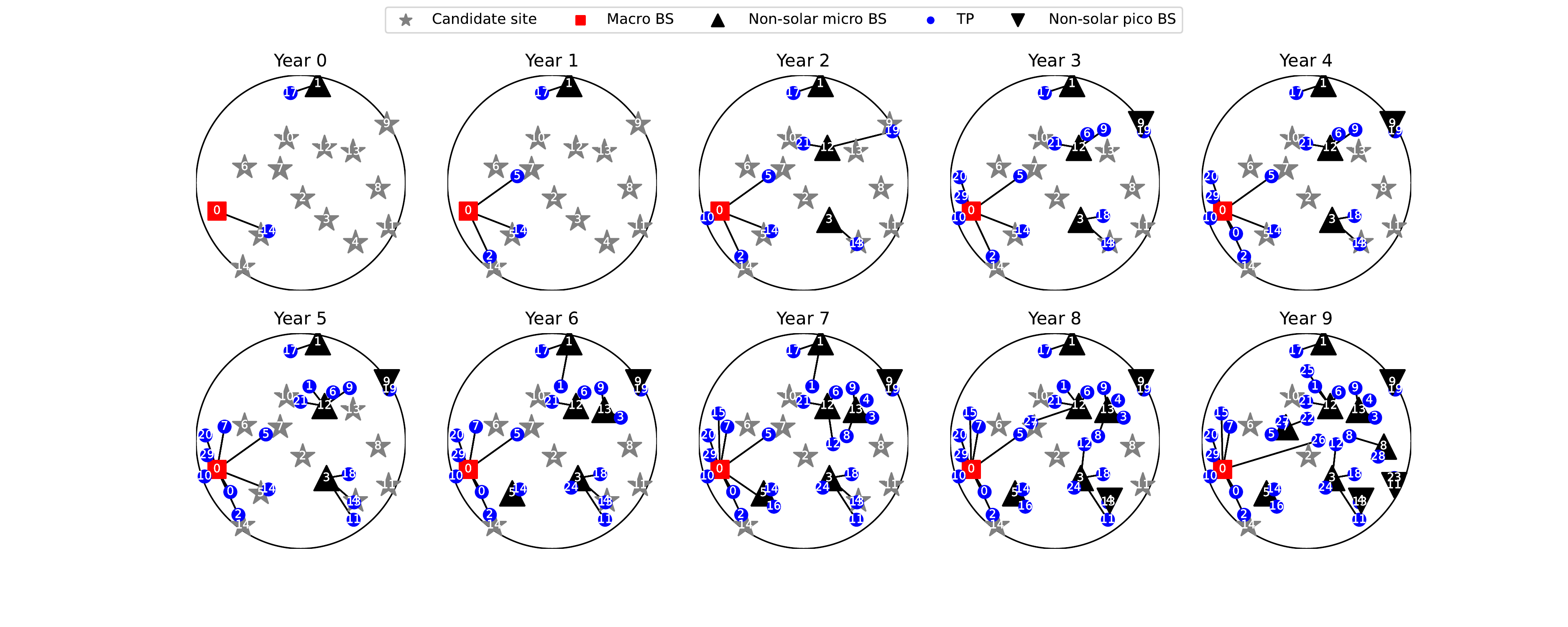}
\caption{Network P2 (Without taxes and Scenario S+Z)}
\label{Network P2(Without taxes and Scenario S+Z)}
\end{figure*}
\subsection{Without taxes}
In this part, we do not consider emissions taxes. Our objective is to
see if free market economics can drive the introduction of  solar BSs
to reduce GHG emissions.  
Figures~\ref{Network P1(Without taxes and Scenario S+Z)} and
\ref{Network P2(Without taxes and Scenario S+Z)} show the base station
installation planning for the S+Z scenario  for P1 and P2 without
taxes. One can also see the assignment of each of the TPs to BSs every
year. The growing demand over the years requires the
installation of new stations. It turns out that the optimal solution
is \emph{not} to install solar base stations both for low and high
traffic.  This is due to the current high 
costs of solar equipment and also to the cost of batteries replacement.
Batteries for green micro BSs have a  lifetime of 7~years 
and 5~years for green pico BSs. 
This is less than the planning horizon so that the batteries have to
be replaced. It then turns out that  with the current  grid prices, the
traditional BSs system is less expensive so that even over a
10-year horizon, the installation of stations that run on solar
energy is not justified.

This can be seen by comparing the results of Table~\ref{Results
  without taxes (Costs)} for the scenarios S+Z and FS+Z. In the second
case, we force the installation of solar equipment on all base
stations. For P1, this reduces the operating cost $Z_o$ from 17.2 to 16.3
with a saving of 0.9 while the capital cost of solar equipment $Z_s$
increases from 21.16 to 23.8 for an added expense of 2.64, 
larger than the savings on the operating cost. The same results can be
found for P2, where the savings is 5.7 and the added cost, 7.4.
The corresponding results of Table~\ref{Results without taxes (Energy)} also show
that forcing the use of solar will have a significant impact on CO2 emissions.

It should be noted that while  solar stations do not emit CO2, there
is still some amount of CO2 emissions  in the Full S + Z scenario.
We  assumed that the system already includes non-solar macro
stations BSs that still produce emissions because of their large
energy use.

If we consider only GHG emissions, PV-Battery BSs, which produce no
GHG, are very attractive as compared with
traditional BSs. When we take into account  realistic cost data, as we do here, the
investment in solar equipment is
not  profitable for cellular operators.  The size of the batteries and
PV needed to reduce the GHG emissions is such that this increases the
price of the solar equipment and thus the total  CAPEX above that for
non-solar networks.

Even if solar base stations are not viable, we can also see from
scenario Z that cell zooming by itself can reduce the  CO2 emissions
and the operating cost, with  a net saving on the total cost.
Thus cell zooming is a good compromise compared with the Full S+Z scenario.

\begin{table}
\centering
\begin{tabular}{lllllll}
  \hline
 & Scnr & $Z$ & $\Delta$(\%) & $Z_s$ & $Z_o$ & $Z_g$ \\
 \hline
\multirow{8}{*}{P1} & B & 39.4 & 0 & 21.16 & 18.3 & 18.3 \\
 & S & 39.4 & 0 & 21.16 & 18.3 & 18.3 \\
 & O & 39.4 & -0.01 & 21.16 & 18.2 & 18.2 \\
 & Z & 38.4 & -2.69 & 21.16 & 17.2 & 17.2 \\
 & S+O & 39.4 & -0.01 & 21.16 & 18.2 & 18.2 \\
 & S+Z & 38.4 & -2.69 & 21.16 & 17.2 & 17.2 \\
 & S+Z0 & 53.1 & 34.83 & 35 & 18.1 & 18.1 \\
 & FS+Z & 40.1 & 1.65 & 23.8 & 16.3 & 16.3 \\
 \hline
\multirow{8}{*}{P2} & B & 77.82 &  0 &  55.85 &  21.96 & 21.96 \\
 & S & 77.82	& 0	& 55.85 & 21.96 & 21.96 \\
 & O & 77.77	& -0.07 & 55.85 & 21.91 & 21.91 \\
 & Z &  74.89	& -3.77 & 55.85 & 19.03 & 19.03 \\
 & S+O &  77.77	& -0.07 & 55.85 & 21.91 & 21.91 \\
 & S+Z & 74.89	& -3.76 & 55.85 & 19.03 & 19.03 \\
 & S+Z0 &  105.96	& 36.16 & 85	& 20.95 & 20.95 \\
 & FS+Z & 79.91	& 2.69 & 63.62 & 16.28 & 16.28 \\

\end{tabular}
\caption{Results without taxes (Costs)}
\label{Results without taxes (Costs)}
\end{table}
\begin{table}
\centering
  \begin{tabular}{lllll}  \hline
     &Scnr & $E_N$ & $E_G$  & CO$_2$ \\
 \hline
\multirow{8}{*}{P1} &B &136 &136 &372 \\
 &S &136 &136 &372 \\
 &O &136 &136 &371 \\
 &Z &126 &126 &346 \\
 &S+O &136 &136 &371 \\
 &S+Z &126 &126 &346 \\
 &S+Z0 &132 &132 &362 \\
 &FS+Z &128 &118 &323 \\
 \hline
\multirow{8}{*}{P2} &B & 164.67 & 164.67 & 451 \\
 &S &164.67 & 164.67 & 451 \\
 &O & 164.21	& 164.21 & 450  \\
 &Z & 140.67	& 140.67 & 385 \\
 &S+O & 164.21	& 164.21 & 450 \\
 &S+Z &140.67	& 140.67 & 385 \\
 &S+Z0 & 153	& 153 & 419 \\
 &FS+Z & 144.91 &117.93 & 323 \\

\end{tabular}
\caption{Results without taxes (Energy).}
\label{Results without taxes (Energy)}
\end{table}

\subsection{With taxes}
 
In this section, we study how a GHG tax can improve the 
the environmental impact of the network.
A GHG tax on fuels is set at a price of \$50 per ton
of ECO2 in the first year and  increased by \$50 each
year until it reaches \$500 in the 10th year.
We consider that coal is the source of electricity and  that 1~kWh of
coal-produced electricity emits~1 kg ECO2. 

We arrived at the tax growth formula by a trial-and-error procedure. We  started
with a constant value of \$50, the carbon tax for
2022~\cite{CarbonTaxCanada} in Canada.  
Initial results showed that this was not large enough to yield
solutions with  green BSs. 
We then tried  small progressive increases from this value, without success.
We finally found  that the carbon
tax had to increase by a factor of ten over the planning horizon to produce
solutions with  green BSs, 
which is an important finding of this project.

\begin{table*}
\centering
  \begin{tabular}{llllllllllll}
    \hline
  Pb & Scnr &  $Z$ & $\Delta$ & $Z_c$ & $Z_s$ & $Z_o$ &
$Z_g$ & $Z_{C02}$ & $Z_s$/kW & $Z_s^\ast$/kW \\
\hline
\multirow{8}{*}{P1} & B & 62.5 & .0 & 21.2 & .0 & 41.3 & 18.3 & 23.1 &
NA & NA \\
 & S & 59.9 & -4.1 & 24.1 & 2.9 & 35.9 & 16.3 & 19.5 & 0.09 & 0.17 \\
 & O & 62.5 & .0 & 21.2 & .0 & 41.3 & 18.2 & 23.1 & NA & NA \\
 & Z & 59.5 & -4.8 & 21.2 & .0 & 38.3 & 17.2 & 21.1 & NA & NA \\
 & S+O & 59.9 & -4.1 & 24.1 & 0 & 35.8 & 16.3 & 19.5 & 0.09 & 0.17 \\
 & S+Z & 58.9 & -5.7 & 23.0 & 1.8 & 36.0 & 16.4 & 19.6 & 0.07 & 0.22 \\
 & S+Z0 & 75.3 & 20.4 & 38.3 & 17.103 & 37.0 & 16.9 & 20.1 & 0.08 & 0.21 \\
 & FS+Z & 59.5 & -4.8 & 23.8 & 2.6 & 35.7 & 16.3 & 19.4 & 0.08 & 0.28
 \\
 \hline
\multirow{8}{*}{P2} & B & 106.15& 0.00& 55.86& 0.0& 50.30&21.96&28.33& NA& NA \\
 & S &  100.84 &-5.00& 63.97 &8.1& 36.87&16.64& 20.23 &0.1 &0.17 \\
 & O &  106.01& -0.14& 55.86 &0.0 &50.15& 21.91&28.24& NA& NA \\
 & Z &  98.69& -7.03 &55.86& 0.0 &42.83&19.03&23.80 &NA &NA\\
 & S+O & 106.01& -0.14 &55.86 &0.0 &50.15&21.91& 28.24& NA & NA\\
 & S+Z & 97.81& -7.86 & 60.97 & 5.1 & 36.84& 16.64& 20.20& 0.1& 0.21 \\
 & S+Z0 & 131.26& 23.66 &91.53& 35.7& 39.73 & 18.08& 21.66& 0.08 &0.22 \\
 & FS+Z & 99.36& -6.40 &63.62& 7.8 & 35.73 & 16.29 & 19.45& 0.08& 0.26 \\

\end{tabular}
\caption{Results with taxes (Costs)}
\label{Results with taxes (Costs)}
\end{table*}

\begin{table}
\centering
\begin{tabular}{lllllllll}
  \hline
  Pb & Scnr & $E_N$ & $E_G$ & CO$_2$ & $E_{si}$ & $E_{su}$ \\
\hline
\multirow{8}{*}{P1} & B & 135.62&135.62& 372&0&0 \\
 & S & 135.62&118.34& 324&31.359&17.278 \\
 & O &  135.57&135.57& 371&0&0 \\
 & Z & 126.14&126.14& 346&0&0 \\
 & S+O & 135.57&135.57& 371&0&0 \\
 & S+Z &  127.43&118.7& 325&26.455&8.739 \\
 & S+Z0 & 136.99&122.11& 335&39.53&14.883 \\
 & FS+Z & 127.82&117.75&323&31.431&10.075 \\

 \hline
\multirow{8}{*}{P2} & B & 164.67& 164.67& 451& 0& 0 \\
 & S & 164.67& 121.55& 333 &71.96 &43.115 \\
 & O &  164.21& 164.21& 450 & 0 &0 \\
 & Z & 140.67& 140.67& 385 &0& 0 \\
 & S+O & 164.21 & 164.21& 450& 0& 0 \\
 & S+Z &  142.52& 121.43& 333& 62.626& 21.091 \\
 & S+Z0 & 159.11& 131.1 & 359& 79.06 & 28.012 \\
 & FS+Z & 145.85& 117.93& 323& 77.622& 27.926 \\

\end{tabular}
\caption{Results with taxes (Energy)}
\label{Results with taxes (Energy)}
\end{table}

Table~\ref{Results with taxes (Costs)} and~\ref{Results with taxes
  (Energy)} list the cost and energy when GHG taxes are considered. We
can see that the total costs are increased  both for P1 and P2
due to the rise of the OPEX. In most scenarios, when solar is enabled,
it is used in the planning. We can also notice a reduction of 47~tons
of C02 emissions for P1 and 118~tons for P2.


\begin{figure*}
\includegraphics[width=\linewidth,height=5cm]{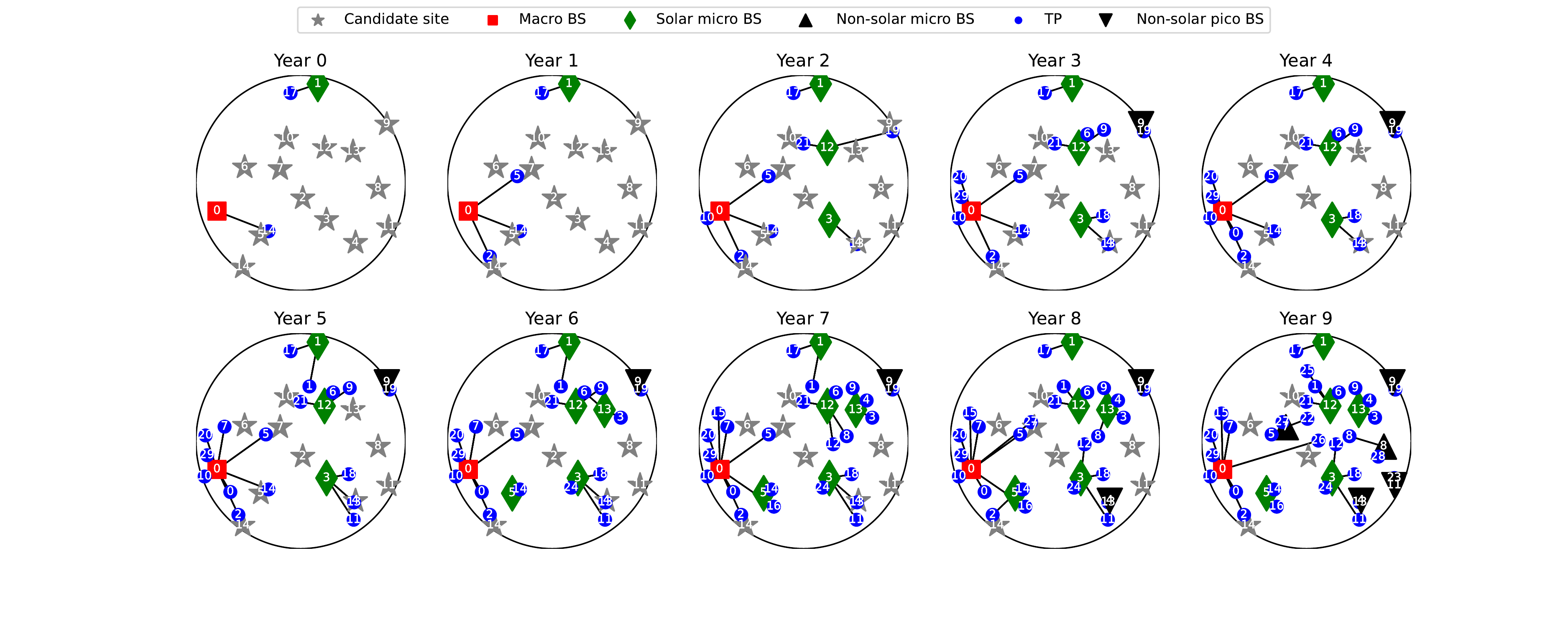}
\caption{Network P2(With taxes and Scenario S)}
\label{Network P2(With taxes and Scenario S)}
\end{figure*}

\begin{figure*}
\includegraphics[width=\linewidth,height=5cm]{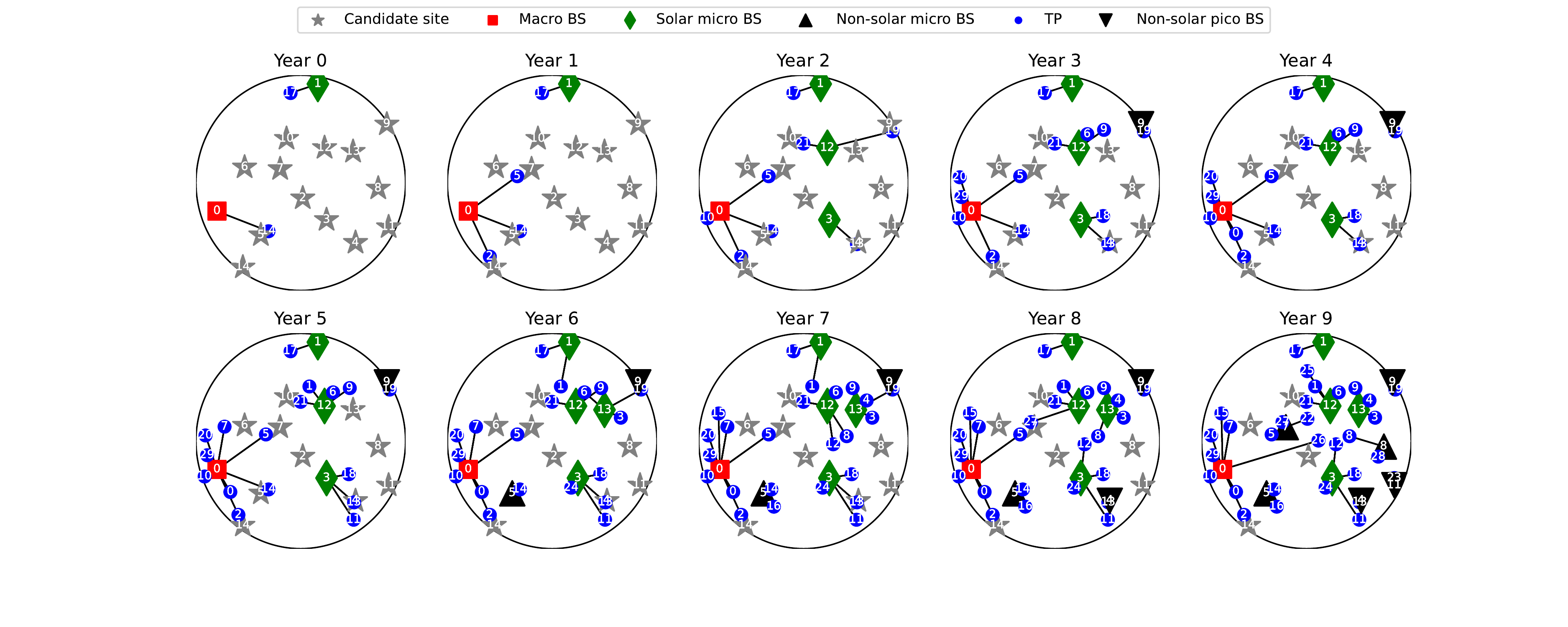}
\caption{Network P2(With taxes and Scenario S+Z)}
\label{Network P2(With taxes and Scenario S+Z)}
\end{figure*}

As we have seen from the cost tables, the largest part of the
total capital cost comes from the installation of the base stations
themselves while the operating cost is only a small fraction of this
total.
We can see an important result from Figures~\ref{Network P2(With taxes
  and Scenario S)} and~\ref{Network   P2(With taxes and Scenario
  S+Z)}.   Both scenarios produce basically the same sequence
of base station installations with the single exception of base
station~9 in year~6. In the S scenario with only the solar option,
the optimal solution is to install a \emph{solar} micro base station. If we
allow cell zooming, however, we need only to install a \emph{non-solar} micro
base station. Because the solar cost is relatively large, this
produces a significant decrease in the capital cost of solar
equipment, from~8.1 to~5.1. In other words, the introduction of cell
zooming can reducee not only the operating cost, as expected, but also
the \emph{capital} cost, in the present case by choosing a less
expensive equipment. It is quite conceivable that in other cases,
this might lead to actually postponing a capital expense with a large
reduction in the total cost. 

\begin{figure*}
\includegraphics[width=\linewidth,height=5cm]{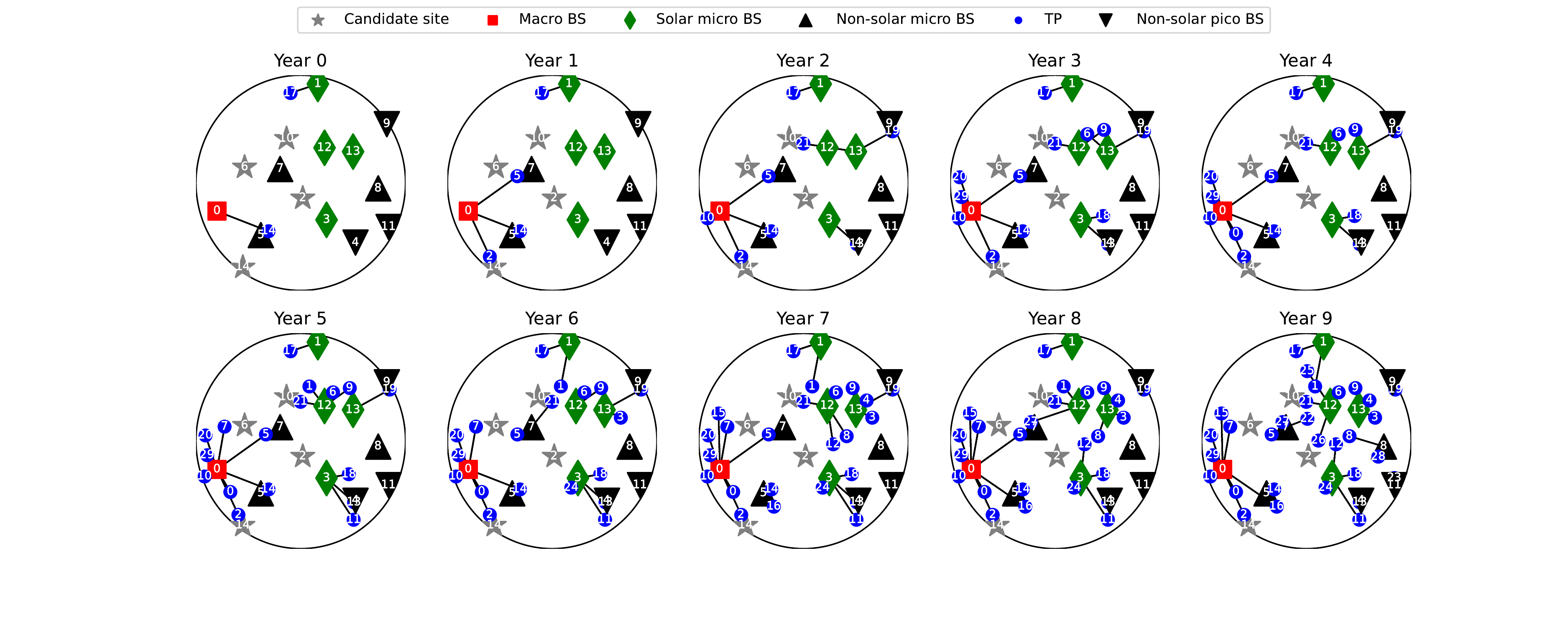}
\caption{Network P2(With taxes and Scenario S + Z  +  Z free at year 0  )}
\label{Network P2(With taxes and Scenario S + Z  +  Z free at year 0  )}
\end{figure*}

We can also see why a static model like S+Z0, where the decision is
taken only at the beginning of the study period, can produce such
costly solutions. From Figure~\ref{Network P2(With taxes and
  Scenario S + Z  +  Z free at year 0  )}, we see that a large number
of base stations  are installed at period~0 but are not used until
(much) later.

\begin{figure*}
\includegraphics[width=\linewidth,height=5cm]{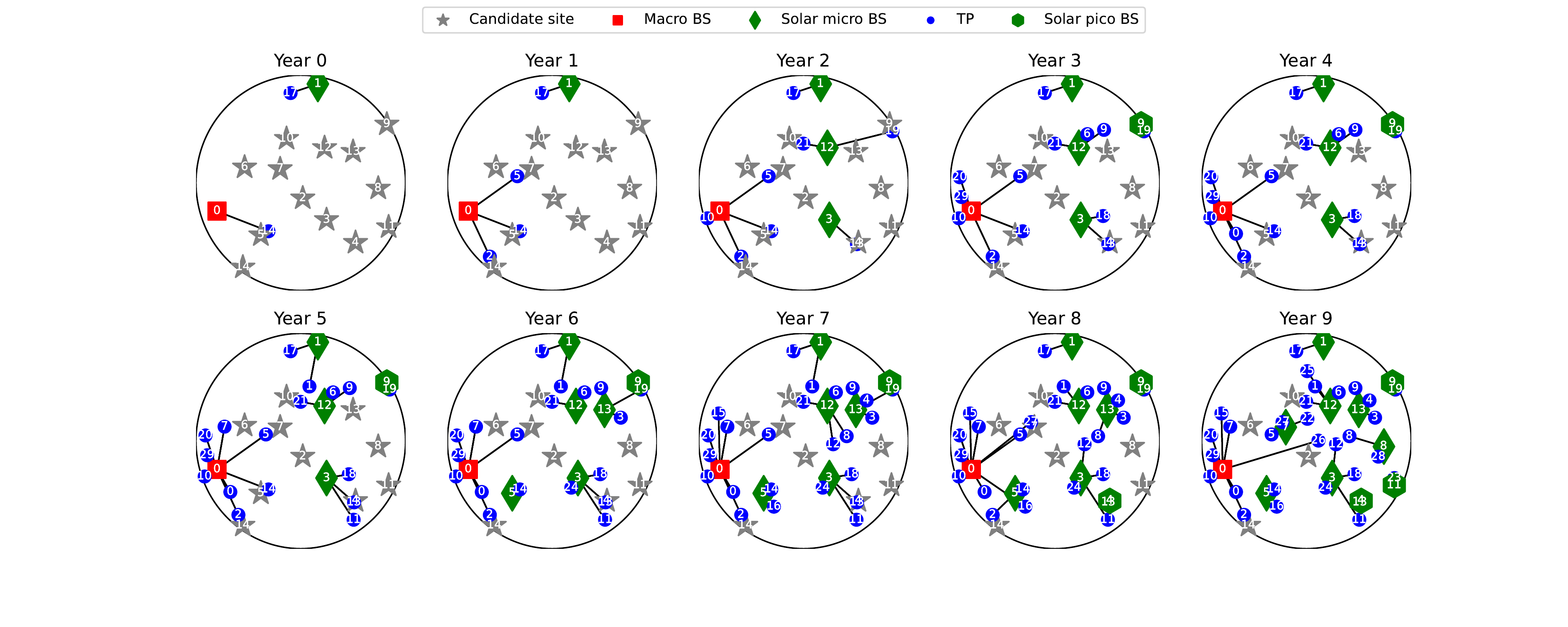}
\caption{Network P1(With taxes and Scenario Full S + Z)}
\label{Network P1(With taxes and Scenario Full S + Z)}
\end{figure*}

We can also see from Figure~\ref{Network P1(With taxes and Scenario Full S + Z)}
why scenario FS+Z, where solar is installed everywhere in network
P2, is more expensive. All the non-solar base stations that were
installed in the S+Z scenario are now forced to be solar which is
where the cost increase comes from.

\subsection{Daily Assignment of TPs to BS}
We can examine in some detail 
the TP assignment to the BS for P1 and the  S+Z scenario.
We see that the dynamic assignment of test points to base stations 
uses  sleep mode in an optimal way
in each period to  reduce both the OPEX and CAPEX. 
\begin{figure}
\centering
\includegraphics[width=\linewidth]{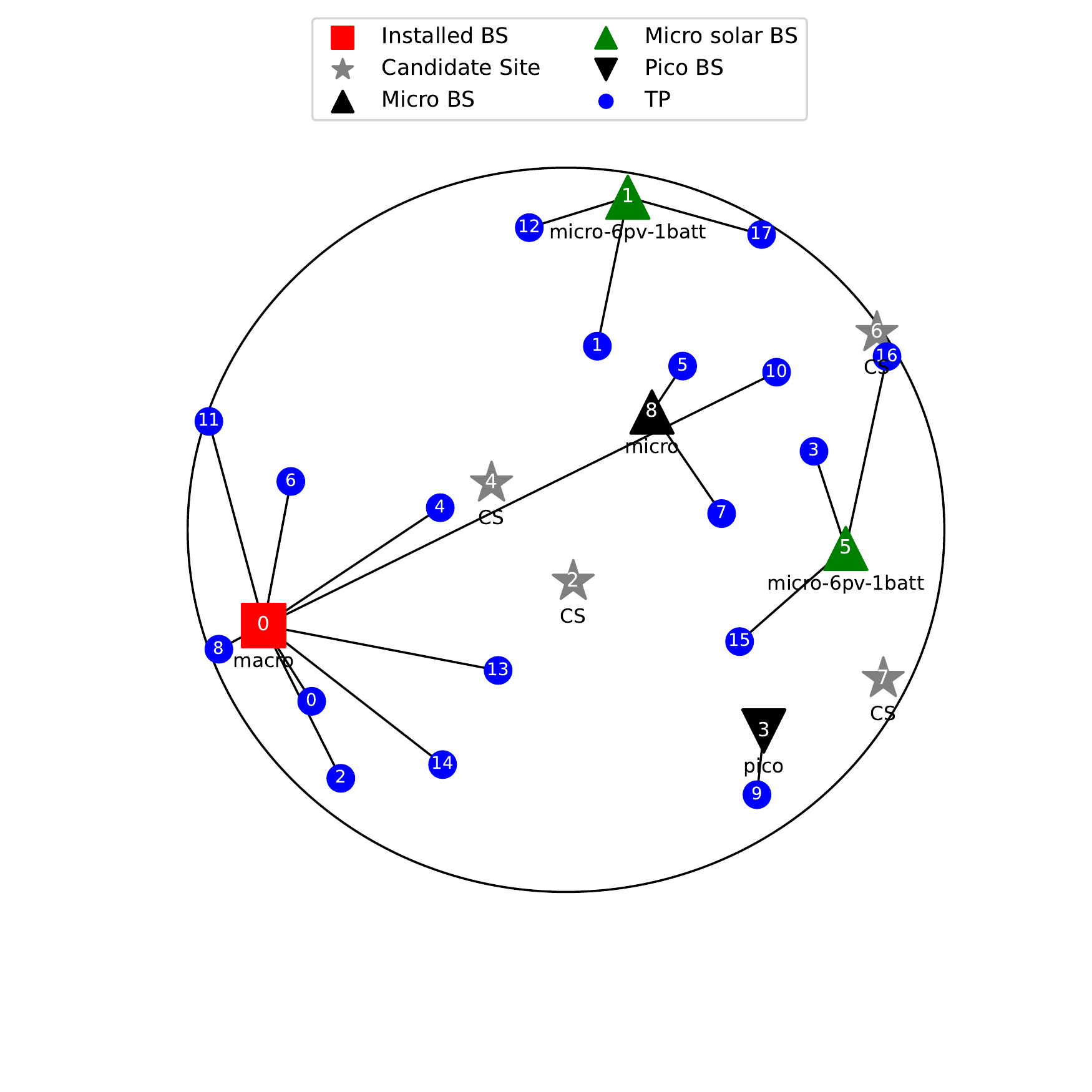}
\caption{Assignment of TPs at 10:00}
\label{assignation_10_sz_p1_true}
\end{figure}
Figure~\ref{assignation_10_sz_p1_true} shows the dynamics of the
network with the peak  traffic demand  at period 10:00.
\begin{figure}
\centering
\includegraphics[width=\linewidth]{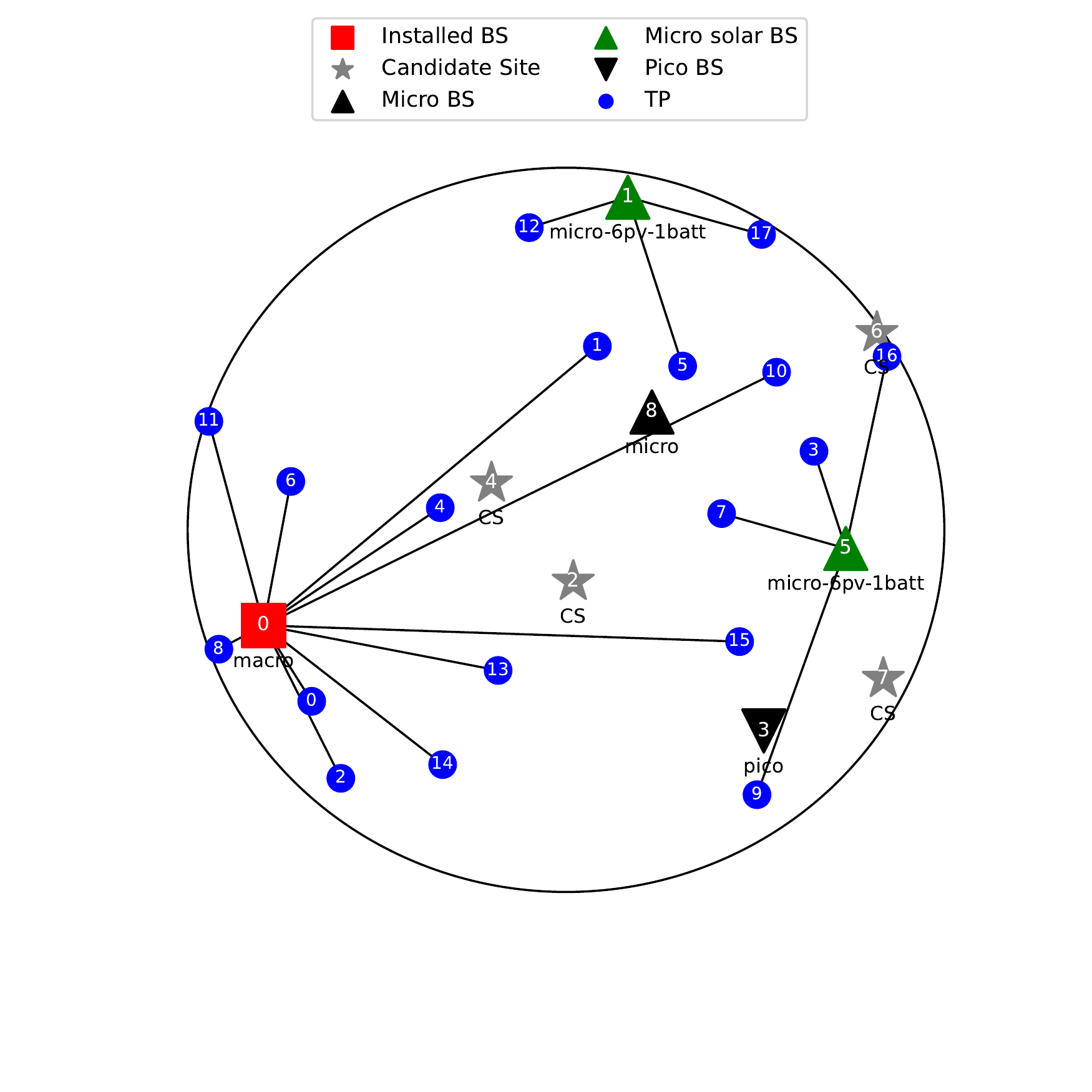}
\caption{Assignment of TPs at 13:00}
\label{assignation_13_sz_p1_true}
\end{figure}
We see on Figure~\ref{assignation_13_sz_p1_true}
that base stations  3 and 5 are in standby mode at 13:00.
\begin{figure}
\centering
\includegraphics[width=\linewidth]{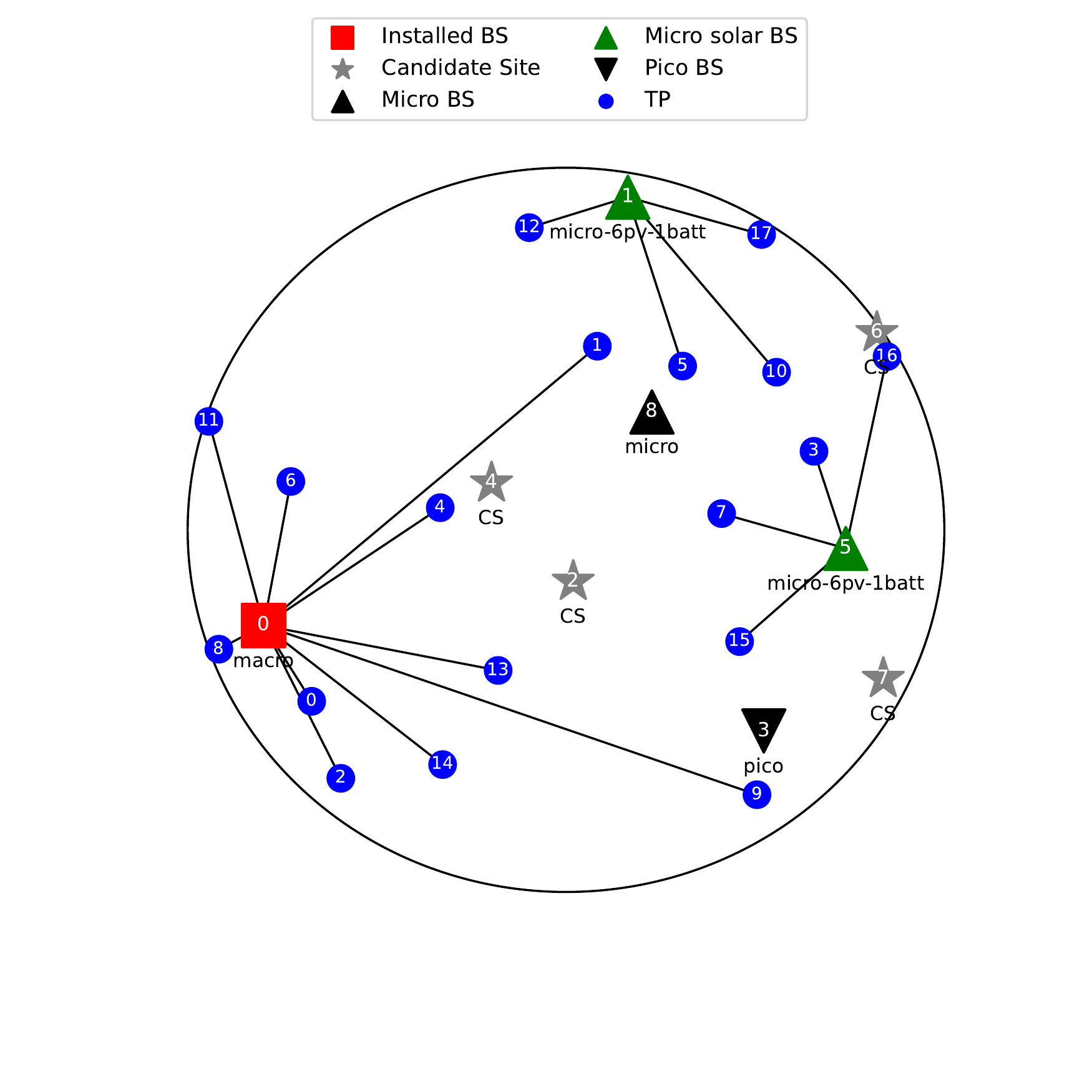}
\caption{Assignment of TPs at 15:00}
\label{assignation_15_sz_p1_true}
\end{figure}
At 15:00, BS 3 is in standby mode in Figure~\ref{assignation_15_sz_p1_true}.

\section{Conclusions}

With the advent of 5G and beyond, there will be a significant increase
in energy use and green gas emissions in cellular
networks. A potential solution is to use solar-powered base stations in the
access in addition to other energy management
techniques.

In this article, we propose a comprehensive
optimization model that not only shows the technical integration of
solar power with other management features, but also allows an
in-depth study of  its economics. The  objective is to reduce the total
cost, including CAPEX and OPEX, over a ten-year horizon and  under
constraints that reflect the inter-relation between networking and
energy. Our aim is  to study the viability of green networks
in reducing costs and GHGs. We use data that is as realistic as
possible to study the technical and economic interactions.

We found
that cellular zooming offers an effective solution for reducing
planning and operating costs in a cellular network. We also found that
cell zooming can not only reduce operating costs but also capital
costs by avoiding the installation of expensive solar stations.
However, in terms
of greening future networks, we found that current energy and
solar-related equipment prices are such that green base stations are not
economically viable for cellular operators.

We then  added a high and
progressive carbon tax  to the  model. This produced networks where the CO2 emissions were
drastically reduced as green base stations were added into the
planning. A conclusion is that unless there is a significant reduction
in the cost of solar equipment, some form of carbon tax will be needed and that that tax has to be significantly higher than current values. 




\appendix

\section{Demand Model}
\label{sec:demandmodel}
The main requirement from users is generally expressed in terms of the
bit rate of a connection. In our model, demand is expressed in terms
of $E^T_{i,j,q,t}$, the energy required by the base station to serve a
user so that the demand in bit rate needs to be converted into an
energy demand. 

This in turn is related to the notion of coverage. In current
networks, where base stations are always turned on at a fixed power,
we can define a fixed coverage radius for a given bit rate to the 
users. With cell zooming and the ensuing dynamic user allocation, the
coverage radius is no longer a fixed base station parameter but now
depends on the transmission power of the base station at some 
instant.

In this section, we provide a simple model for this extended coverage
radius concept and also a transformation rule from a demand in bit
rates into an energy demand. This model is more realistic than the one
used in~\cite{damours18} where these values were generated randomly. 
\subsection{Coverage}
\label{sec:coverage}
In order to model the coverage, we assume and additive, white noise
channel. This could easily extended to more complex models if
needed. In that case, the maximum bit rate $\overline{\rho}_{i,j,l,s}$
that a test point $i$ can receive from a base station $j$ of type $l$
transmitting at the power level of state $s$ is given by
\begin{equation} \label{eq:shannon}
\overline{\rho}_{i,j,l,s} =  B_{l} \log_2 \left( 1 + \frac{W^r_{i,j,l,s}}{N} \right) 
\end{equation}
where
\begin{description}
\item $B_l$ The bandwidth of the channels used by a base station of type $l$
\item $W^r_{i,j,l,s}$ The power received by the test point
\item $N$ The channel noise.
\end{description}
A base station can meet a required bit rate $\rho_{i,q,t}$ from a test
point $i$ at time $t$ from year $q$ if $\rho_{i,q,t} \leq 
\overline{\rho}_{i,j,l,s}$.
From this, we can define the generalized coverage parameter
$\vect{k}$ for each time instant by the condition. 
\begin{align}
  \label{eq:calculk}
k_{i,j,l,s,q,t} & = \begin{cases}
1 & \text{if $\rho_{i,q,t} \le \overline{\rho}_{i,j,l,s}$}\\
0 & \text{otherwise}
\end{cases}\\
\forall\,
i \in \mathcal{I},\ &
\ j \in \mathcal{B  \cup C},
\ l \in \mathcal{L}, \ s \in \mathcal{S}_l ,
\ q \in \mathcal{Q} ,
\ t \in \mathcal{T} .
\nonumber 
\end{align}
\subsubsection{Channel Gain}
\label{sec:chgain}
The signal power received by a test point from a base station
transmitting at a given power is given by
\begin{align}
	W^r_{i,j,l,s} &=W^{x}_{l,s} \gamma_{i,j} \label{eq:wrwx} \\
	\gamma_{i,j} &= \frac{\alpha}{d_{i,j}^n} \label{eq:airgain}
      \end{align}
where
\begin{description}
\item[$W^x_{i,j,l,s}$]~ is the base station transmission power
\item[$W^r_{i,j,l,s}$]~ is the power received at the test point
\item[$d_{i,j}$] the distance between $i$ and $j$
\item[$n$] the propagation coefficient. 
\item[$\alpha$] the antenna gain
\end{description}
\subsubsection{Energy Demand}
\label{sec:enerdem}
We can now compute the power required from a base station by a test
point by replacing~\eqref{eq:wrwx} and~\eqref{eq:airgain} 
into~\eqref{eq:shannon} and solving for $W^x_{i,j,q,t}$ to get 
\begin{equation} \label{eq:transreqpw}
W^{x}_{i,j,q,t} = \frac{N d_{i,j}^n}{\alpha} \left(
2^{ \rho_{i,q,t} / B_j} -1  \right)
\end{equation}
so that the energy requirement during period $(q,t)$ is given by
\begin{equation} \label{eq:transreqenergy}
E^T_{i,j,q,t} = \Delta t W^{x}_{i,j,q,t}
\end{equation}

\bibliography{telecom}

\begin{thebibliography}{10}
\providecommand{\url}[1]{#1}
\csname url@samestyle\endcsname
\providecommand{\newblock}{\relax}
\providecommand{\bibinfo}[2]{#2}
\providecommand{\BIBentrySTDinterwordspacing}{\spaceskip=0pt\relax}
\providecommand{\BIBentryALTinterwordstretchfactor}{4}
\providecommand{\BIBentryALTinterwordspacing}{\spaceskip=\fontdimen2\font plus
\BIBentryALTinterwordstretchfactor\fontdimen3\font minus
  \fontdimen4\font\relax}
\providecommand{\BIBforeignlanguage}[2]{{%
\expandafter\ifx\csname l@#1\endcsname\relax
\typeout{** WARNING: IEEEtran.bst: No hyphenation pattern has been}%
\typeout{** loaded for the language `#1'. Using the pattern for}%
\typeout{** the default language instead.}%
\else
\language=\csname l@#1\endcsname
\fi
#2}}
\providecommand{\BIBdecl}{\relax}
\BIBdecl

\bibitem{bohli19}
A.~Bohli and R.~Bouallegue, ``How to meet increased capacities by future green
  {5G} networks: A survey,'' \emph{IEEE Access}, vol.~7, pp. 42\,220--42\,237,
  2019.

\bibitem{chamola16}
V.~Chamola and B.Sikdar, ``Solar powered cellular base stations:current
  scenario, issues and proposed solutions,'' \emph{{IEEE} Communications
  Magazine}, pp. 108--114, may 2016.

\bibitem{malmodin18}
\BIBentryALTinterwordspacing
J.~Malmodin and D.~Lund\'en, ``The energy and carbon footprint of the global
  ict and e\&m sectors 2010–2015,'' \emph{Sustainability}, vol.~10, no.~9,
  2018. [Online]. Available: \url{https://www.mdpi.com/2071-1050/10/9/3027}
\BIBentrySTDinterwordspacing

\bibitem{freitag21a}
\BIBentryALTinterwordspacing
C.~Freitag, M.~Berners-Lee, K.~Widdicks, B.~Knowles, G.~S. Blair, and
  A.~Friday, ``The real climate and transformative impact of ict: A critique of
  estimates, trends, and regulations,'' \emph{Patterns}, vol.~2, no.~9, p.
  100340, 2021. [Online]. Available:
  \url{https://www.sciencedirect.com/science/article/pii/S2666389921001884}
\BIBentrySTDinterwordspacing

\bibitem{israr11}
\BIBentryALTinterwordspacing
A.~Israr, Q.~Yang, W.~Li, and A.~Y. Zomaya, ``Renewable energy powered
  sustainable 5g network infrastructure: Opportunities, challenges and
  perspectives,'' \emph{Journal of Network and Computer Applications}, vol.
  175, p. 102910, 2021. [Online]. Available:
  \url{https://www.sciencedirect.com/science/article/pii/S1084804520303702}
\BIBentrySTDinterwordspacing

\bibitem{williams22}
\BIBentryALTinterwordspacing
L.~Williams, B.~K. Sovacool, and T.~J. Foxon, ``The energy use implications of
  5g: Reviewing whole network operational energy, embodied energy, and indirect
  effects,'' \emph{Renewable and Sustainable Energy Reviews}, vol. 157, p.
  112033, 2022. [Online]. Available:
  \url{https://www.sciencedirect.com/science/article/pii/S1364032121012958}
\BIBentrySTDinterwordspacing

\bibitem{wu15b}
G.~Wu, C.~Yang, S.~Li, and G.~Y. Li, ``Recent advances in energy-efficient
  networks and their application in 5g systems,'' \emph{IEEE Wireless
  Communications}, vol.~22, no.~2, pp. 145--151, 2015.

\bibitem{ancans17}
\BIBentryALTinterwordspacing
G.~Ancans, V.~Bobrovs, A.~Ancans, and D.~Kalibatiene, ``Spectrum considerations
  for {5G} mobile communication systems,'' \emph{Procedia Computer Science},
  vol. 104, pp. 509--516, 2017, iCTE 2016, Riga Technical University, Latvia.
  [Online]. Available:
  \url{https://www.sciencedirect.com/science/article/pii/S1877050917301679}
\BIBentrySTDinterwordspacing

\bibitem{hu20}
S.~Hu, X.~Chen, W.~Ni, X.~Wang, and E.~Hossain, ``Modeling and analysis of
  energy harvesting and smart grid-powered wireless communication networks: A
  contemporary survey,'' \emph{IEEE Transactions on Green Communications and
  Networking}, vol.~4, no.~2, pp. 461--496, 2020.

\bibitem{wikipedia_cost_electricity}
\BIBentryALTinterwordspacing
``Cost of electricity by source.'' [Online]. Available:
  \url{https://en.wikipedia.org/wiki/Cost_of_electricity_by_source}
\BIBentrySTDinterwordspacing

\bibitem{alsharif15}
\BIBentryALTinterwordspacing
M.~H. Alsharif, R.~Nordin, and M.~Ismail, ``Energy optimisation of hybrid
  off-grid system for remote telecommunication base station deployment in
  {Malaysia},'' \emph{EURASIP Journal on Wireless Communications and
  Networking}, vol. 2015, no.~1, p.~64, Mar 2015. [Online]. Available:
  \url{https://doi.org/10.1186/s13638-015-0284-7}
\BIBentrySTDinterwordspacing

\bibitem{olubayo19}
\BIBentryALTinterwordspacing
O.~M. Babatunde, I.~H. Denwigwe, D.~E. Babatunde, A.~O. Ayeni, T.~B. Adedoja,
  and O.~S. Adedoja, ``Techno-economic assessment of photovoltaic-diesel
  generator-battery energy system for base transceiver stations loads in
  {Nigeria},'' \emph{Cogent Engineering}, vol.~6, no.~1, p. 1684805, 2019.
  [Online]. Available: \url{https://doi.org/10.1080/23311916.2019.1684805}
\BIBentrySTDinterwordspacing

\bibitem{flavio21}
\BIBentryALTinterwordspacing
F.~Odoi-Yorke and A.~Woenagnon, ``Techno-economic assessment of solar pv/fuel
  cell hybrid power system for telecom base stations in {Ghana},'' \emph{Cogent
  Engineering}, vol.~8, no.~1, p. 1911285, 2021. [Online]. Available:
  \url{https://doi.org/10.1080/23311916.2021.1911285}
\BIBentrySTDinterwordspacing

\bibitem{aderemi18}
\BIBentryALTinterwordspacing
B.~A. Aderemi, S.~P.~D. Chowdhury, T.~O. Olwal, and A.~M. Abu-Mahfouz,
  ``Techno-economic feasibility of hybrid solar photovoltaic and battery energy
  storage power system for a mobile cellular base station in {Soshanguve},
  {South} {Africa},'' \emph{Energies}, vol.~11, no.~6, 2018. [Online].
  Available: \url{https://www.mdpi.com/1996-1073/11/6/1572}
\BIBentrySTDinterwordspacing

\bibitem{baidas21}
\BIBentryALTinterwordspacing
M.~W. Baidas, R.~W. Hasaneya, R.~M. Kamel, and S.~S. Alanzi, ``Solar-powered
  cellular base stations in {Kuwait}: A case study,'' \emph{Energies}, vol.~14,
  no.~22, 2021. [Online]. Available:
  \url{https://www.mdpi.com/1996-1073/14/22/7494}
\BIBentrySTDinterwordspacing

\bibitem{alsharif21b}
M.~H. Alsharif, R.~Kannadasan, A.~Jahid, M.~A. Albreem, J.~Nebhen, and B.~J.
  Choi, ``Long-term techno-economic analysis of sustainable and zero grid
  cellular base station,'' \emph{IEEE Access}, vol.~9, pp. 54\,159--54\,172,
  2021.

\bibitem{Ibrahim21}
\BIBentryALTinterwordspacing
I.~A. Ibrahim, S.~Sabah, R.~Abbas, M.~Hossain, and H.~Fahed, ``A novel sizing
  method of a standalone photovoltaic system for powering a mobile network base
  station using a multi-objective wind driven optimization algorithm,''
  \emph{Energy Conversion and Management}, vol. 238, p. 114179, 2021. [Online].
  Available:
  \url{https://www.sciencedirect.com/science/article/pii/S0196890421003551}
\BIBentrySTDinterwordspacing

\bibitem{niyato12}
D.~Niyato, X.~Lu, and P.~Nanyang, ``Adaptive power management for wireless base
  stations in a smart grid environment,'' \emph{IEEE Wireless Communications},
  vol.~19, no.~6, pp. 44--51, Dec. 2012.

\bibitem{liu14}
L.~Liu, S.~Men, M.~Liu, and B.~Zhou, ``An energy saving solution for wireless
  communication equipment,'' in \emph{IEEE 36th International
  Telecommunications Energy Conference (INTELEC)}, 2014, pp. 1--3.

\bibitem{alonso18}
\BIBentryALTinterwordspacing
S.~Herrería-Alonso, M.~Rodríguez-Pérez, M.~Fernández-Veiga, and
  C.~López-García, ``An optimal dynamic sleeping control policy for single
  base stations in green cellular networks,'' \emph{Journal of Network and
  Computer Applications}, vol. 116, pp. 86--94, 2018. [Online]. Available:
  \url{https://www.sciencedirect.com/science/article/pii/S1084804518301760}
\BIBentrySTDinterwordspacing

\bibitem{boiardi12}
\BIBentryALTinterwordspacing
S.~Boiardi, A.~Capone, and B.~Sansó, ``Joint design and management of
  energy-aware mesh networks,'' \emph{Ad Hoc Networks}, vol.~10, no.~7, pp.
  1482--1496, 2012. [Online]. Available:
  \url{https://www.sciencedirect.com/science/article/pii/S1570870512000765}
\BIBentrySTDinterwordspacing

\bibitem{mollahasani19}
S.~Mollahasani and E.~Onur, ``Density-aware, energy- and spectrum-efficient
  small cell scheduling,'' \emph{IEEE Access}, vol.~7, pp. 65\,852--65\,869,
  2019.

\bibitem{jahid20}
A.~Jahid, M.~Hossain, M.~Monju, M.~Rahman, and M.~Hossain, ``Techno-economic
  and energy efficiency analysis of optimal power supply solutions for green
  cellular base stations,'' \emph{IEEE Access}, vol.~8, pp. 43\,776--43\,795,
  2020.

\bibitem{xu18}
X.~Xu, C.~Yuan, W.~Chen, X.~Tao, and Y.~Sun, ``Adaptive cell zooming and
  sleeping for green heterogeneous ultradense networks,'' \emph{IEEE
  Transactions on Vehicular Technology}, vol.~67, no.~2, pp. 1612--1621, 2018.

\bibitem{tipper10}
D.~Tipper, A.~Rezgui, P.~Krishnamurthy, and P.~Pacharintanakul, ``Dimming
  cellular networks,'' in \emph{{IEEE} GLOBECOM}, 2010, pp. 1--6.

\bibitem{budzisz14}
L.~Budzisz, F.~Ganji, G.~Rizzo, M.~A. Marsan, M.~Meo, Y.~Zhang, G.~Koutitas,
  L.~Tassiulas, S.~Lambert, B.~Lannoo, M.~Pickavet, A.~Conte, I.~Haratcherev,
  and A.~Wolisz, ``Dynamic resource provisioning for energy efficiency in
  wireless access networks: A survey and an outlook,'' \emph{{IEEE}
  Communication Surveys \& Tutorials}, vol.~16, no.~4, pp. 2259--2285, 2014.

\bibitem{vereecken11}
W.~Vereecken, W.~V. Heddeghem, M.~Deruyck, B.~Puype, B.~Lannoo, W.~Joseph,
  D.~Colle, L.~Martens, and P.~Demeester, ``Power consumption in
  telecommunication networks: overview and reduction strategies,''
  \emph{Communications Magazine}, vol.~49, no.~6, pp. 62--69, 2011.

\bibitem{wu15}
J.~Wu, Y.~Zhang, M.~Zukerman, and E.~K.-N. Yung, ``Energy-efficient
  base-stations sleep-mode techniques in green cellular networks: A survey,''
  \emph{Communication Surveys \& Tutorials}, vol.~17, no.~2, pp. 803--826,
  2015.

\bibitem{wu13}
Y.~Wu, H.~Gaoning, S.~Zhang, Y.~Chen, and S.~Xu, ``Energy efficient coverage
  planning in cellular networks with sleep mode,'' in \emph{Proc. {IEEE} 24th
  International Symposium on Personal, Indoor and Mobile Radio Communications},
  Sep. 2013, pp. 2586--2590.

\bibitem{tsilimantos13}
\BIBentryALTinterwordspacing
D.~Tsilimantos, J.~Gorce, and E.~Altman, ``Stochastic analysis of energy
  savings with sleep mode in {OFDMA} wireless networks,'' in \emph{Proceedings
  IEEE INFOCOM}.\hskip 1em plus 0.5em minus 0.4em\relax IEEE, 2013, pp.
  1097--1105. [Online]. Available:
  \url{https://dx.doi.org/10.1109/INFCOM.2013.6566900}
\BIBentrySTDinterwordspacing

\bibitem{chiaraviglio08b}
L.~Chiaraviglio, D.~Ciullo, M.~Meo, and M.~Marsan, ``Energy-aware {UMTS} access
  networks,'' in \emph{WPMC'08}, 2008.

\bibitem{niu10}
Z.~Niu, Y.~Wu, J.~Gong, and Z.~Yang, ``Cell zooming for cost-efficient green
  cellular networks,'' \emph{Communications Magazine}, vol.~48, no.~11, pp.
  74--79, Nov. 2010.

\bibitem{yigitel14}
\BIBentryALTinterwordspacing
M.~Yigitel, O.~Incel, and C.~Ersoy, ``Dynamic base station planning with power
  adaptation for green wireless cellular networks,'' \emph{Eurasip Journal on
  Wireless Communications and Networking}, vol. 2014, no.~1, p.~77, 2014.
  [Online]. Available: \url{https://dx.doi.org/10.1186/1687-1499-2014-77}
\BIBentrySTDinterwordspacing

\bibitem{wang17}
\BIBentryALTinterwordspacing
B.~Wang, Q.~Yang, L.~T. Yang, and C.~Zhu, ``On minimizing energy consumption
  cost in green heterogeneous wireless networks,'' \emph{Computer Networks},
  vol. 129, pp. 522--535, Dec. 2017. [Online]. Available:
  \url{https://doi.org/10.1016/j.comnet.2017.03.024}
\BIBentrySTDinterwordspacing

\bibitem{boiardi13}
S.~Boiardi, A.~Capone, and B.~Sans\`o, ``Radio planning of energy-aware
  cellular networks,'' \emph{Computer Networks}, vol.~57, pp. 2564--2577, 2013.

\bibitem{boiardi14}
------, ``Planning for energy-aware wireless networks,'' \emph{IEEE
  Communications Magazine}, vol.~52, no.~2, pp. 156--162, Feb. 2014.

\bibitem{damours18}
M.~D'Amours, A.~Girard, and B.~Sansò, ``Planning solar in energy-managed
  cellular networks,'' \emph{{IEEE} Access}, vol.~6, pp. 65\,212--65\,226, Oct.
  2018.

\bibitem{arnold10}
O.~Arnold, F.~Richter, G.~Fettweis, and O.~Blume, ``Power consumption modeling
  of different base station types in heterogeneous cellular networks,'' in
  \emph{Future Network Mobile Summit}, 2010, pp. 1--8.

\bibitem{claussen10}
\BIBentryALTinterwordspacing
H.~Claussen, I.~Ashraf, and L.~T.~W. Ho, ``Dynamic idle mode procedures for
  femtocells,'' \emph{Bell Labs Technical Journal}, vol.~15, no.~2, pp.
  95--116, 2010. [Online]. Available:
  \url{https://onlinelibrary.wiley.com/doi/abs/10.1002/bltj.20443}
\BIBentrySTDinterwordspacing

\bibitem{zhang17}
Y.~Zhang, M.~Meo, R.~Gerboni, and M.~A. Marsan, ``Minimum cost solar power
  systems for {LTE} macro base stations,'' \emph{Computer Networks}, vol. 112,
  pp. 12--23, Oct. 2017.

\bibitem{balakrishnan21}
A.~Balakrishnan, S.~De, and L.-C. Wang, ``Network operator revenue maximization
  in dual powered green cellular networks,'' \emph{IEEE Transactions on Green
  Communications and Networking}, vol.~5, no.~4, pp. 1791--1805, 2021.

\bibitem{zheng13}
M.~Zheng, P.~Pawelczak, S.~Stanczak, and H.~Yu, ``Planning of cellular networks
  enhanced by energy harvesting,'' \emph{IEEE Communications Letters}, vol.~17,
  no.~6, pp. 1092--1095, 2013.

\bibitem{chen18}
Y.~Chen, L.~Duan, and Q.~Zhang, ``Financial analysis of network upgrade,''
  \emph{IEEE Transactions on Vehicular Technology}, vol.~67, no.~6, pp.
  5496--5499, 2018.

\bibitem{xu15}
J.~Xu, L.~Duan, and R.~Zhang, ``Cost-aware green cellular networks with energy
  and communication cooperation,'' \emph{{IEEE} Communications Magazine}, pp.
  257--263, may 2015.

\bibitem{johansson04}
\BIBentryALTinterwordspacing
K.~Johansson, A.~Furuskar, P.~Karlsson, and J.~Zander, ``Relation between base
  station characteristics and cost structure in cellular systems,'' in
  \emph{15th International Symposium on Personal, Indoor and Mobile Radio
  Communications ({PIMRC})}, Barcelona, Spain, Sep. 2004, pp. 2627--2631.
  [Online]. Available:
  \url{https://ieeexplore.ieee.org/abstract/document/1368795}
\BIBentrySTDinterwordspacing

\bibitem{eia-solar-cost21}
\BIBentryALTinterwordspacing
``Cost and performance characteristics of new generating technologies, annual
  energy outlook 2021,'' Feb. 2021. [Online]. Available:
  \url{https://www.eia.gov/outlooks/aeo/assumptions/pdf/table_8.2.pdf}
\BIBentrySTDinterwordspacing

\bibitem{marsan13}
M.~Marsan, G.~Bucalo, A.~Di~Caro, M.~Meo, and Y.~Zhang, ``Towards zero grid
  electricity networking: Powering {BS}s with renewable energy sources,'' in
  \emph{IEEE International Conference on Communications Workshops}, Jun. 2013,
  pp. 596--601.

\bibitem{energyhub21}
\BIBentryALTinterwordspacing
R.~Urban, ``Electricity prices in {Canada},'' 2021. [Online]. Available:
  \url{https://www.energyhub.org/electricity-prices/}
\BIBentrySTDinterwordspacing

\bibitem{HydroQuebecInstaller}
\BIBentryALTinterwordspacing
{Hydro-Québec}. (2019) Costs and affordability: Don’t be blinded by the
  (sun)light! [Online]. Available:
  \url{https://www.hydroquebec.com/solar/costs.html}
\BIBentrySTDinterwordspacing

\bibitem{ouarzazate22}
\BIBentryALTinterwordspacing
``Rayonnement solaire à {Ouarzazate}.'' [Online]. Available:
  \url{https://fr.tutiempo.net/radiation-solaire/ouarzazate.html}
\BIBentrySTDinterwordspacing

\bibitem{CarbonTaxCanada}
\BIBentryALTinterwordspacing
{Ministry of Environment and Climate Change Strategy}, ``{British Columbia}'s
  carbon tax,'' 2022. [Online]. Available:
  \url{https://www2.gov.bc.ca/gov/content/environment/climate-change/clean-economy#carbontax}
\BIBentrySTDinterwordspacing

\end{thebibliography}
\bibliographystyle{ieeetran}
\end{document}